\documentclass[pdflatex,sn-nature]{sn-jnl}

\usepackage{graphicx}
\usepackage{amsmath}
\usepackage{amssymb}
\usepackage{xcolor}
\usepackage{url}

\newcommand{\BGPLogGain}{1.97\%}
\newcommand{\BGPLogCI}{[0.23\%, 3.37\%]}

\newcommand{\BGPLogContentGain}{2.58\%}
\newcommand{\BGPLogContentCI}{[2.17\%, 3.09\%]}
\newcommand{\BGPLogInventoryGain}{0.96\%}

\newcommand{\BGPLogTimePositive}{7}

\newcommand{\BGPHgbGain}{9.54\%}
\newcommand{\BGPHgbCI}{[7.18\%, 11.45\%]}

\newcommand{\BGPHgbContentGain}{1.20\%}
\newcommand{\BGPHgbContentCI}{[0.36\%, 2.25\%]}
\newcommand{\BGPHgbInventoryGain}{-0.32\%}

\newcommand{\BGPHgbTimePositive}{10}

\begin{document}

\title[Causal-fate dynamics]{Causal-fate dynamics of unrealized influence}

\author[1]{\fnm{Yiwei} \sur{Liu}}\email{yiweiliu1@link.cuhk.edu.cn}
\author*[2]{\fnm{Luwei} \sur{Yang}}\email{yangluwei@sribd.cn}
\author*[1,2]{\fnm{Shunbo} \sur{Lei}}\email{leishunbo@cuhk.edu.cn}

\affil[1]{\orgdiv{School of Science and Engineering}, \orgname{The Chinese University of Hong Kong, Shenzhen}, \orgaddress{\state{Guangdong}, \postcode{518172}, \country{China}}}
\affil[2]{\orgname{Shenzhen Research Institute of Big Data (SRIBD)}, \orgaddress{\city{Shenzhen}, \postcode{518172}, \country{China}}}

\abstract{Many dynamical systems generate influences whose consequences are not fully exhausted in the realized trajectory at the moment they arise. Such consequences are often treated as absent, delayed or statically stored, leaving unclear how unrealized influence retains future relevance as the system evolves. Here we formulate causal-fate dynamics, in which generated influence may be realized, remain latent, or be transformed by subsequent dynamics, and give an exact finite-transport representation when the relevant maps are specified. A connectome-constrained \textit{Caenorhabditis elegans} model first motivates the biological hypothesis that unresolved inter-neuronal influence may persist and contribute to later propagation; it does not establish such a mechanism in living animals. We next examine operational Internet routing, where a dynamically updated cross-observer history retains predictive information beyond the current local route state. We then use the representation to construct a Transformer architecture that explicitly transports and selectively realizes latent contextual influence while retaining language-modeling function. The three studies distinguish a model-motivated scientific hypothesis, an observational phenomenon compatible with future-relevant history and an executable construction for carrying unrealized influence through subsequent computation.}

\maketitle

\section{Introduction}\label{sec:introduction}

Dynamical systems are typically described in terms of what has already been realized: forces that have acted, interactions that have occurred, decisions that have been executed and exchanges that have been completed~\cite{sontag1998mathematical,wiggins2003nonlinear}. This realized-state perspective underlies state-space and realization theory~\cite{kalman1960filtering,hokalman1966realization}, as well as operator-based descriptions of dynamical evolution~\cite{koopman1931hamiltonian,mezic2005spectral}. It is complete, nevertheless, only when the present realized state contains every consequence of the system's past that can still shape its future.

Many systems do not satisfy this condition. Network constraints~\cite{hespanha2007networked}, selective interactions~\cite{heemels2012event}, activation barriers~\cite{kramers1940brownian,hanggi1990reaction}, competing objectives~\cite{boyd2004convex,rawlings2017model} and distributed coordination can separate what a system generates from what enters its realized trajectory. This separation cannot, in general, be reduced to delay alone. A generated influence may be partially expressed, redistributed, transformed, cancelled or absorbed, while its unfinished consequence may later re-enter the realized trajectory in a different form, or may never enter it at all. The central question is therefore what becomes of a consequence that is absent from the realized present but not yet exhausted from future evolution.

What persists across time is not necessarily the unrealized action itself, nor a fixed backlog awaiting eventual release. Rather, it is the still-unexhausted consequence of non-realization. We formalize this idea as causal-fate dynamics. In this representation, an influence may be generated with a definite counterfactual consequence, partially realized in the present, retained as a latent consequence, transported by subsequent dynamics and transformed before it re-enters the realized trajectory, disappears or assumes a different functional role. Generation, realization, latency, transport and transformation thus provide a minimal language for a form of history that is absent from the present trajectory yet remains consequential for future evolution.

This perspective changes the meaning of state in systems in which influence can remain unfinished. The realized state alone need not determine the future if two histories have generated different latent consequences that have not yet been exhausted. The dynamically relevant state must then include not only what has occurred, but also what has been generated, withheld and transformed by the dynamics through which it has passed. Multi-channel and multi-unit systems provide concrete settings in which individual units continue through local evolution while non-local influence remains latent, is reshaped over time and may later enter the realized trajectory. Such systems thereby instantiate causal-fate dynamics in systems with interacting components.

We investigate causal-fate dynamics through three complementary studies, moving from scientific hypothesis through observational compatibility to explicit construction. First, a connectome-constrained \textit{Caenorhabditis elegans} model tests whether selective realization can preserve a propagation structure informed by the public signal-propagation atlas~\cite{randi2023neural,wormneuroatlas2023}. This model-based result motivates a biological hypothesis about carriers of unresolved inter-neuronal influence. Next, in Border Gateway Protocol (BGP) routing, a dynamically updated state formed from unmatched cross-observer updates retains information about future local route changes beyond that contained in the current local route state~\cite{rekhter2006bgp,routeviews2025,riperis2025}. This observational phenomenon is compatible with future-relevant history under the adopted representation. Finally, we construct a Transformer in which token-local evolution, latent contextual influence and selective realization form an executable architecture~\cite{vaswani2017attention,allal2025smollm2,merity2016pointer}.

The three studies have distinct evidential roles. The neural model does not establish a biological mechanism, and the routing observations do not verify the exact finite-transport equations. The Transformer implements those equations by construction. Their exactness follows from the specified maps and recurrence; the scientific questions concern which representations are physically meaningful, predictively useful or capable of guiding functional design.

\section{Results}\label{sec:results}

\subsection{Unrealized influence becomes a dynamical state}

\begin{figure*}[t]
    \centering
    \includegraphics[width=\textwidth]{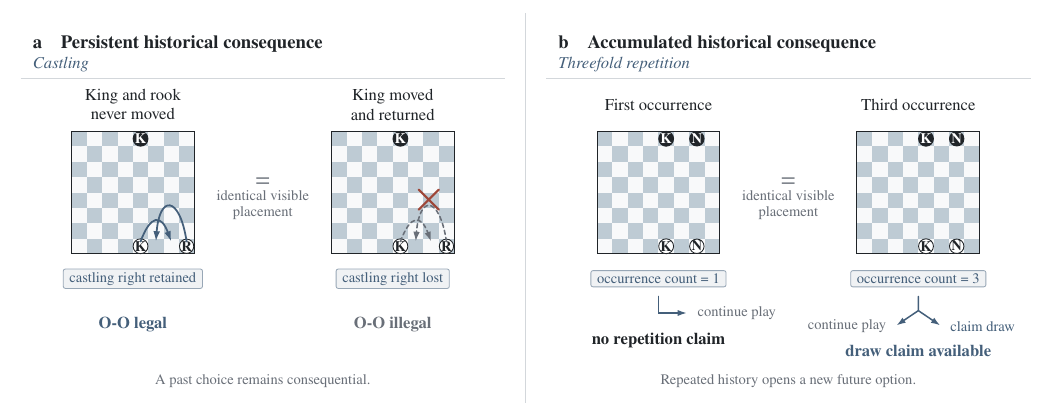}
    \caption{\textbf{Identical visible chess configurations can admit different futures.}
    \textbf{a, Persistent historical consequence.} The visible piece placement and side to move are identical, but kingside castling is legal only when neither the king nor the rook has previously moved. Moving either piece and returning it to its original square does not restore the lost castling right.
    \textbf{b, Accumulated historical consequence.} The visible placement, side to move and move rights are identical, but a draw claim becomes available only when the same position has occurred for at least the third time.
    These rule-defined examples show that an identical visible configuration alone need not contain all history-dependent information relevant to future admissibility.}
    \label{fig:chess-history-witness}
\end{figure*}

We first establish how unrealized influence can be incorporated into the
dynamical state. If an influence generated at one step need not be fully
expressed at that same step, the realized state alone may fail to contain every
consequence of the system's past that remains relevant to its future
evolution. The missing information cannot be represented by an unconstrained
record of earlier events. It must instead be carried by a state whose evolution
tracks the still-unexhausted consequence of influence that has been generated but not realized. A rule-defined example offers an intuitive illustration of this distinction: in chess, identical
visible piece configurations can retain different history-dependent rights or
claims and therefore admit different futures
(Fig.~\ref{fig:chess-history-witness}).

Let \(x_t\in\mathcal X\) denote the realized state at evolution step \(t\),
where \(\mathcal X\) is the realized-state space and \(t\) may represent physical time, a decision stage, an iterative
update or computational depth. Let \(\mathcal C\) denote the set of influence
channels. For each channel \(c\in\mathcal C\), let \(\mathcal U_c\) denote its influence space, and let
\(\phi_{c,t}(x)\in\mathcal U_c\) be the influence newly generated through that
channel when the current state is \(x\). Let
\(\psi_{c,t}\in\mathcal U_c\) denote the net contribution through channel \(c\)
that is actually admitted to the realized update at step \(t\). The distinction between these quantities is fundamental:
\(\phi_{c,t}(x_t)\) describes what the current realized state generates, whereas
\(\psi_{c,t}\) describes what the realized trajectory receives at the current step.
When previously unrealized influence is allowed to re-enter the trajectory, \(\psi_{c,t}\)
may include influence generated at the current step, influence transported
from earlier steps, or both. It is therefore not, in
general, a predetermined part of \(\phi_{c,t}(x_t)\).

Let \(\mathcal L_t:\mathcal X\rightarrow\mathcal X\) describe the evolution produced by the existing realized state and
the system's internal or local
dynamics before any channel-level influence is admitted at step \(t\). Let
\(\mathcal F_t:\mathcal X\rightarrow\mathcal X\) denote the corresponding fully
realized map, and let \(x_t^{\mathrm F}\) denote its fully realized reference
trajectory. If all newly generated influence were admitted immediately and in
full, the reference trajectory would satisfy
\begin{equation}
x_{t+1}^{\mathrm F}
=
\mathcal F_t(x_t^{\mathrm F}),
\qquad
\mathcal F_t(x)
=
\mathcal L_t(x)
+
\sum_{c\in\mathcal C}
B_c\phi_{c,t}(x),
\label{eq:fully-realized-map}
\end{equation}
where \(B_c:\mathcal U_c\rightarrow\mathcal X\) embeds the influence associated with channel \(c\) into the realized-state coordinates. The trajectory \(x_t^{\mathrm F}\) is a constitutive counterfactual reference: it is the
trajectory obtained when every influence generated at a given step is admitted
to the realized state at that same step.

Throughout this work, the term \emph{causal} is used relative to a
specified intervention on realization. The fully realized reference is
obtained by replacing the actual realization mechanism with immediate
admission of every generated influence, while holding the finite system
maps, local dynamics, exogenous inputs and initial condition fixed.
The resulting difference therefore represents a constitutive intervention
contrast defined within the dynamical model. The fully realized reference serves as
the counterfactual comparator, whereas the
realized trajectory may admit a different channel-level contribution:
\begin{equation}
x_{t+1}
=
\mathcal L_t(x_t)
+
\sum_{c\in\mathcal C}
B_c\psi_{c,t}.
\label{eq:realized-map}
\end{equation}

To represent the consequence that remains outside the realized trajectory,
associate each channel \(c\) with a latent influence state
\(\xi_{c,t}\in\mathcal U_c\), and let
\(
\boldsymbol{\xi}_t
=
(\xi_{c,t})_{c\in\mathcal C}
\)
collect the channel-level latent states. Let \(\mathcal T_{c,t}\) denote the
channel-\(c\) transport operator. Its output
\(\mathcal T_{c,t}(x_t,\boldsymbol{\xi}_t)\in\mathcal U_c\) represents the contribution assigned to channel \(c\) after the previously latent influence has been propagated through the intervening dynamics.
The total influence available through that channel at step \(t\) is therefore
\begin{equation}
h_{c,t}
=
\mathcal T_{c,t}(x_t,\boldsymbol{\xi}_t)
+
\phi_{c,t}(x_t).
\label{eq:channel-available-influence}
\end{equation}
After the contribution \(\psi_{c,t}\) has been admitted to the realized
update, the remainder becomes the next latent state:
\begin{equation}
\xi_{c,t+1}
=
h_{c,t}-\psi_{c,t}
=
\mathcal T_{c,t}(x_t,\boldsymbol{\xi}_t)
+
\phi_{c,t}(x_t)
-
\psi_{c,t}.
\label{eq:latent-state}
\end{equation}

The recurrence in Eq.~\eqref{eq:latent-state} is a channel-level balance
relation. The transport operator \(\mathcal T_{c,t}\) need not
preserve the magnitude, direction or functional interpretation of the
previously latent influence. It may absorb, redirect or transform historical
influence, so its output is not necessarily an unchanged copy of a past action
but the current form assigned to an unfinished consequence after the
surrounding system has evolved. Accordingly, \(\xi_{c,t}\) is not merely a
record that an influence was missed in the past; it is the present
channel-level representation of a consequence that has not yet been exhausted.

The dynamically relevant state of the resulting representation, denoted
\(y_t\), is the column concatenation of the realized and latent components:
\begin{equation}
y_t
=
\operatorname{col}
\left(
x_t,
\boldsymbol{\xi}_t
\right).
\label{eq:extended-state}
\end{equation}
where \(\operatorname{col}\) concatenates its arguments.
Two augmented states may share the same realized component \(x_t\) while
differing in their latent components. Under the same subsequent realization
mechanism, they may therefore generate different future realized trajectories. Retaining only \(x_t\)
would leave this dependence implicit; including the latent states \(\boldsymbol{\xi}_t\) makes the
consequences of previous non-realization explicit and carries them forward as
part of the dynamical description.

The recurrence in Eq.~\eqref{eq:latent-state} also yields an exact balance
identity.
Adding and subtracting \(\xi_{c,t}\) gives
\begin{equation}
\phi_{c,t}(x_t)
+
\left[
\mathcal T_{c,t}(x_t,\boldsymbol{\xi}_t)
-
\xi_{c,t}
\right]
=
\psi_{c,t}
+
\left(
\xi_{c,t+1}-\xi_{c,t}
\right).
\label{eq:fate-balance}
\end{equation}
The left-hand side combines the influence newly generated at the current step
with the change induced by transporting the existing latent state. The
right-hand side separates the influence admitted to the realized trajectory
from the net change retained in the latent state. This identity follows
directly from the latent-state recurrence; it is not a predefined partition
of the newly generated influence. The vector balance in
Eq.~\eqref{eq:fate-balance} is not a decomposition into non-negative or mutually
exclusive fractions. The quantities involved are generally vector-valued, and contributions in different
directions may cancel. Moreover, \(\psi_{c,t}\) may contain
influence generated at earlier steps, whose magnitude, direction or functional meaning may have been altered by transport. The identity therefore balances current generation, transport-induced change,
current realization and the net change in latency.

Eqs.~\eqref{eq:fully-realized-map}--\eqref{eq:fate-balance}
establish the channel-level causal-fate representation. Generation,
realization, latency, transport and transformation describe how influence
absent from the realized present can nevertheless remain part of the system's dynamical
history.

\subsection{Finite transport gives latent influence a counterfactual meaning}

It should be noted that the channel-level latent states introduced above would have limited
scientific content if the transport operators governing their evolution could be chosen independently
of the system dynamics. We therefore constrain latent evolution by the
finite system map itself. After the channel-level latent consequences in the realized-state coordinates are embedded and collected, let
\(\bar\xi_t\in\mathcal X\) denote the resulting aggregate latent contribution of all underlying channels.
We assume that \(\mathcal X\) is represented in additive coordinates, so
that \(x_t+\bar\xi_t\) is well defined.

Define the complete state represented by the realized--latent pair as
\begin{equation}
z_t
=
x_t+\bar\xi_t.
\label{eq:complete-state}
\end{equation}
Thus,
\(z_t\) combines the currently realized component \(x_t\) with the aggregate
latent displacement \(\bar\xi_t\).

The fully realized and local maps in Eq.~\eqref{eq:fully-realized-map} first
determine the aggregate influence newly generated by the current realized state:
\begin{equation}
g_t
=
\mathcal F_t(x_t)-\mathcal L_t(x_t)
=
\sum_{c\in\mathcal C}
B_c\phi_{c,t}(x_t).
\label{eq:aggregate-generation}
\end{equation}
Previously unrealized influence need not be carried forward unchanged. After the existing latent displacement has been propagated through the finite system map,
its contribution to the next-state evolution is
\begin{equation}
\tau_t
=
\mathcal F_t(x_t+\bar\xi_t)-\mathcal F_t(x_t).
\label{eq:finite-transport}
\end{equation}
The total influence currently available for realization is therefore
\begin{align}
h_t
&=
g_t+\tau_t
=
\mathcal F_t(x_t+\bar\xi_t)-\mathcal L_t(x_t).
\label{eq:available-influence}
\end{align}
The relation in Eq.~\eqref{eq:available-influence} does not introduce two independently prescribed components. It follows by combining the influence newly
generated by the realized state with the finite effect of transporting the
existing latent displacement through \(\mathcal F_t\).

Let
\begin{equation}
\psi_t
=
\mathcal R_t(h_t)
\label{eq:aggregate-realization}
\end{equation}
denote the aggregate contribution admitted to the realized trajectory at step
\(t\), where \(\mathcal R_t\) is the realization rule. The rule may be
deterministic, state-dependent, thresholded, learned or imposed by the underlying
realization mechanism. Any dependence on the realized state, latent state or
external conditions is understood to be incorporated into \(\mathcal R_t\). When the
channel decomposition is retained, \(\psi_t\) corresponds to the embedded
aggregate of the channel-level realized contributions.

The realized state then evolves according to
\begin{equation}
x_{t+1}
=
\mathcal L_t(x_t)+\psi_t,
\label{eq:realized-update}
\end{equation}
whereas the contribution not admitted at the current step becomes the next
aggregate latent displacement:
\begin{equation}
\bar\xi_{t+1}
=
h_t-\psi_t.
\label{eq:aggregate-latent-update}
\end{equation}
Adding Eqs.~\eqref{eq:realized-update} and
\eqref{eq:aggregate-latent-update}, and substituting
Eq.~\eqref{eq:available-influence}, yields
\begin{equation}
z_{t+1}
=
x_{t+1}+\bar\xi_{t+1}
=
\mathcal F_t(x_t+\bar\xi_t)
=
\mathcal F_t(z_t).
\label{eq:exact-closure}
\end{equation}
The complete represented state therefore evolves exactly under the fully
realized map.

Recall the fully realized counterfactual trajectory \(x_t^{\mathrm F}\)
defined in Eq.~\eqref{eq:fully-realized-map}. If the complete represented
state and the fully realized reference are initialized identically,
\begin{equation}
z_0=x_0^{\mathrm F},
\end{equation}
then Eq.~\eqref{eq:exact-closure} implies
\begin{equation}
z_t=x_t^{\mathrm F}
\qquad
\text{for all }t.
\label{eq:counterfactual-equivalence}
\end{equation}
Consequently,
\begin{equation}
\bar\xi_t
=
x_t^{\mathrm F}-x_t.
\label{eq:latent-gap}
\end{equation}
The aggregate latent state therefore has a direct counterfactual meaning: in
the adopted coordinates, it is precisely the finite displacement that completes the realized state into the fully realized reference state. Once the
fully realized reference and realized trajectory are fixed, the aggregate
latent displacement is uniquely determined by Eq.~\eqref{eq:latent-gap}. Its decomposition into channel-level
latent states may nevertheless be non-unique when the channel embeddings overlap. The formal
statement and proof are given in Supplementary Note 1.

This construction also clarifies why the realized state need not be predictively sufficient.
If two admissible histories reach the same realized state but produce different
futures under the same subsequent realization protocol, then the realized state alone cannot contain all dynamically relevant information. Any exact predictive state must retain additional
history-dependent information. In causal-fate dynamics, this distinction is
represented by the latent state; Supplementary Note 1 gives the corresponding formal result
and its coordinate interpretation.

The finite-transport relation in Eq.~\eqref{eq:finite-transport} further specifies
how an existing latent displacement is propagated. When the fully realized map is affine,
with \(M_t\) its time-indexed linear map and \(b_t\) its offset,
\begin{equation}
\mathcal F_t(x)=M_t x+b_t,
\end{equation}
finite transport is exact and state independent:
\begin{equation}
\mathcal F_t(x_t+\bar\xi_t)-\mathcal F_t(x_t)
=
M_t\bar\xi_t.
\label{eq:affine-transport}
\end{equation}

The three studies below move from a biological hypothesis to a compatible observational phenomenon and then to an explicit construction. In a connectome-constrained neural model, we ask whether selective realization can preserve propagation structure, motivating a hypothesis about unresolved influence in living systems. We next test whether a dynamically updated routing history retains predictive information beyond the current local state. Finally, we implement finite transport and selective realization directly in a Transformer. These studies address different claims: functional possibility in a neural model, observational compatibility in routing and architectural viability in an artificial system.

In the neural and Transformer constructions, the fully realized reference admits every newly generated non-local influence immediately and in full, the local-autonomy counterfactual withholds that influence, and the selective trajectory transports and selectively admits the consequence of non-realization. These comparisons distinguish the need for non-local influence from the need to admit it at every update. The routing analysis does not require the observed system to satisfy this additive construction.

\subsection{Selective inter-neuronal realization sustains propagation in a connectome-constrained neural model}

We first examined whether causal-fate dynamics could support a non-trivial
function in a connectome-constrained natural-system model. The continuous
substrate follows the graded-potential model structure
developed by Kunert, Shlizerman and Kutz~\cite{kunert2014low}, in which membrane voltage and
synaptic-gating variables evolve under leak, external stimulus,
chemical-synaptic and gap-junction currents. The model
contains 188 identified neurons from the \textit{Caenorhabditis elegans} head network and retains both directed chemical synapses and
electrical gap junctions, using the parameter values and connectome
scaling adopted in this study. Experimental data from the neural
signal-propagation atlas of Randi et al.~\cite{randi2023neural} were used to
freeze the functional object, fit the observation readout and evaluate signal
propagation in the connectome-constrained model.

Propagation was evaluated over a common 10-s response window using a common signed-mean observation operator and 1,141 high-confidence stimulus--response pairs. The full empirical propagation object was adopted to instantiate the fixed continuous substrate and the study-specific readout. The 168 stimulus neurons represented in the formal propagation object were then divided once into 118 selection stimuli and 50 evaluation stimuli. This split separated realization operating-point selection from stimulus-stratified evaluation after the continuous model and readout had been fixed. The realization interval and threshold were selected on the 118 selection stimuli and frozen before scoring the remaining 50 evaluation stimuli; the complete protocol is given in Supplementary Note~2.

We applied the general finite-transport construction using the fully coupled neural map and a local-autonomy map that retained intrinsic neuronal dynamics, local gating and external stimulation while withholding newly generated chemical-synaptic and electrical coupling influence. In the selective trajectory, unresolved inter-neuronal influence remained latent, was transported and transformed through the nonlinear neural dynamics, and re-entered the realized trajectory at selected realization steps. This construction preserved the exact represented-state closure; the complete recurrence, realization rule and numerical protocol are given in Supplementary Note~2.

\begin{figure*}[t]
    \centering
    \includegraphics[width=\textwidth]{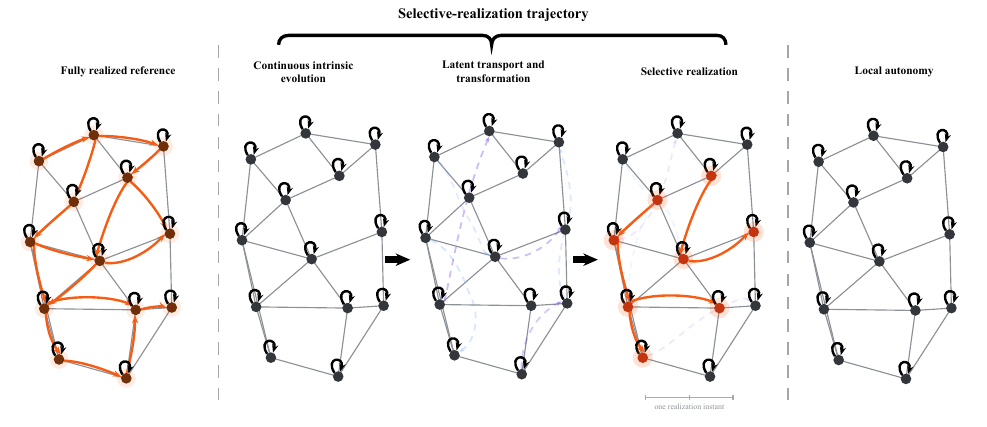}
    \caption{\textbf{Selective realization of inter-neuronal influence in a connectome-constrained \textit{Caenorhabditis elegans} model.} From left to right, the fully realized reference immediately admits all generated cross-neuronal influence. In the selective-realization trajectory, intrinsic neuronal dynamics continue while unrealized influence is transported and transformed before selected components re-enter the realized trajectory. The local-autonomy counterfactual retains intrinsic neuronal evolution but excludes latent transport and subsequent cross-neuronal realization.}
    \label{fig:celegans-propagation}
\end{figure*}

The candidate mechanism and functional result are summarized in Fig.~\ref{fig:celegans-propagation}. Under local-autonomy dynamics, the selected cross-neuronal propagation structure was not reproduced, yielding a relative propagation error of \(1\). By contrast, the frozen selective-realization trajectory preserved the principal propagation structure with a relative error of \(3.51499\%\), a near-perfect Pearson correlation coefficient \((r>0.99)\) and complete sign agreement.

Thus, within the adopted nonlinear neural model, causal-fate dynamics gives an
exact extended-state representation and a functionally non-trivial
selective-realization witness. The result shows that the principal propagation structure can be preserved even when inter-neuronal influence is not admitted immediately and in full at every update. It therefore motivates a concrete biological hypothesis: living neural systems may contain distributed physical carriers of unresolved inter-neuronal influence, whose subsequent transport, release or absorption contributes to neural propagation. The physical identity of such carriers in living animals remains unknown and requires direct experimental investigation. The hypothesis could be tested by perturbing candidate carriers while controlling the measured realized neural state and subsequent input, and then determining whether later responses change in the predicted manner.

\subsection{Ordered operational routing history retains future information beyond the current local state}

BGP announcements and withdrawals appear asynchronously at public routing collectors, including RouteViews Chile and RIPE RIS rrc06~\cite{rekhter2006bgp,routeviews2025,riperis2025}. With respect to a given
collector--peer--prefix tuple, the current local route, its path attributes and the
most recent local update form a strong representation of the visible state. Recent updates already
observed at the other collector but not yet matched by a corresponding local event may nevertheless
form an unfinished propagation front. We tested whether this structured cross-observer history retained information, beyond the current local state, about whether the local peer--prefix route would change within the next $30$ s.

\begin{figure*}[t]
    \centering
    \includegraphics[width=\textwidth]{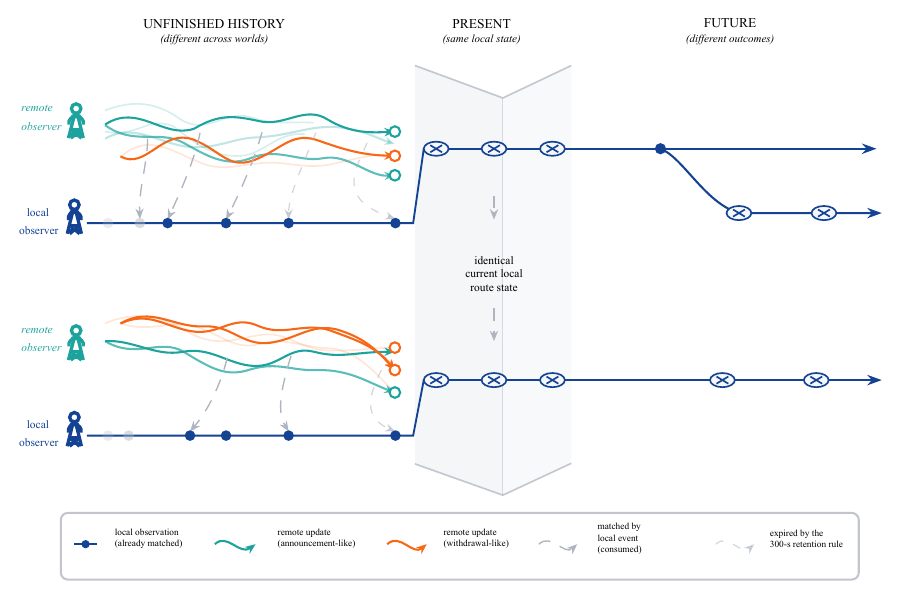}
    \caption{\textbf{Different unfinished cross-observer histories distinguish futures from the same current local route state.} Remote updates not yet matched by local observations form an ordered propagation front that evolves through addition, local matching and 300-s expiry, so identical local presents can retain different information about subsequent route evolution.}
    \label{fig:bgp-history-state}
\end{figure*}

The strong current-state representation captured the current local route, its path attributes and the latest local update. The candidate representation augmented this state with the five newest unmatched cross-observer updates and fixed-dimensional summaries of the eligible unmatched queue. Each local event consumed the oldest matching remote occurrence, and entries older than 300 s expired; simultaneous observations were processed as one group before reading either front. The five-slot view was recomputed after these operations, so consumed or expired events were no longer represented. This bounded readout and its queue summaries formed a finite, ordered and dynamically updated state (Fig.~\ref{fig:bgp-history-state}).

Across held-out data, the corrected five-event unmatched-front representation gave relative log-loss gains of \BGPLogGain{} for Logistic regression and \BGPHgbGain{} for HistGradientBoosting over the strong current-state model, with paired 95\% time-block bootstrap intervals of \BGPLogCI{} and \BGPHgbCI{}, respectively. Gains were positive in \BGPLogTimePositive{} of ten and \BGPHgbTimePositive{} of ten consecutive held-out time blocks.

We separated the role of chronological content from that of slot encoding. Permuting complete event tuples within each sample retained their ages and therefore tested encoding sensitivity, since those ages still revealed chronology. In a separate paired comparison without per-event ages, chronological event content gave gains of \BGPLogContentGain{} and \BGPHgbContentGain{} over within-sample shuffled content, with intervals of \BGPLogContentCI{} and \BGPHgbContentCI{}. Against a canonical unordered inventory of the same event content, the corresponding gains were \BGPLogInventoryGain{} and \BGPHgbInventoryGain{}. The HGB comparison with the canonical inventory had an interval crossing zero, and one HGB seed also gave an interval crossing zero against shuffled content. Chronological arrangement therefore contributed under some tested comparisons, without establishing a uniform advantage over every same-content representation. The complete controls, negative results and seed analyses are reported in Supplementary Note~3.

Reviewed trajectories exhibited both enhancement and consumption or expiry of the
propagation front, demonstrating that the state evolved with subsequent routing
events rather than merely accumulating a static record of the past.

The routing analysis therefore identifies a finite, dynamically updated
history representation with predictive information beyond the tested strong
current-state representation. Its entries change through observed matching and
the specified retention rule. These results support the operational relevance
of future-relevant history; they do not identify the additive latent state of
the finite-transport construction. The next study specifies and implements that construction directly.

\subsection{Causal-fate dynamics defines a viable Transformer architecture}

We next moved from observational history to explicit design, asking whether causal-fate dynamics could serve as an architectural principle. We reorganized the pretrained Transformer so that each token maintained a realized hidden state together with a latent contextual state. The complete pretrained layer transported and transformed the combined state, whereas a token-local counterpart preserved local layer computation without introducing newly generated cross-token influence. A token-level realization rule determined when accumulated contextual influence entered the realized hidden state. The construction preserved exact layerwise closure; the complete recurrence, realization score and threshold-selection protocol are given in Supplementary Note~4.

We instantiated the construction in the 30-layer SmolLM2-135M~\cite{allal2025smollm2} model without modifying its pretrained weights, and evaluated 8,255 effective next-token prediction positions from WikiText-2~\cite{merity2016pointer}.
The fully realized Transformer achieved a test perplexity of $38.6512$. Removing newly generated cross-token influence while retaining token-local computation increased perplexity to \(1315.2426\), demonstrating that cross-token contextual influence is indispensable for the language-modeling function examined here.

\begin{figure*}[t]
    \centering
    \includegraphics[width=\textwidth]{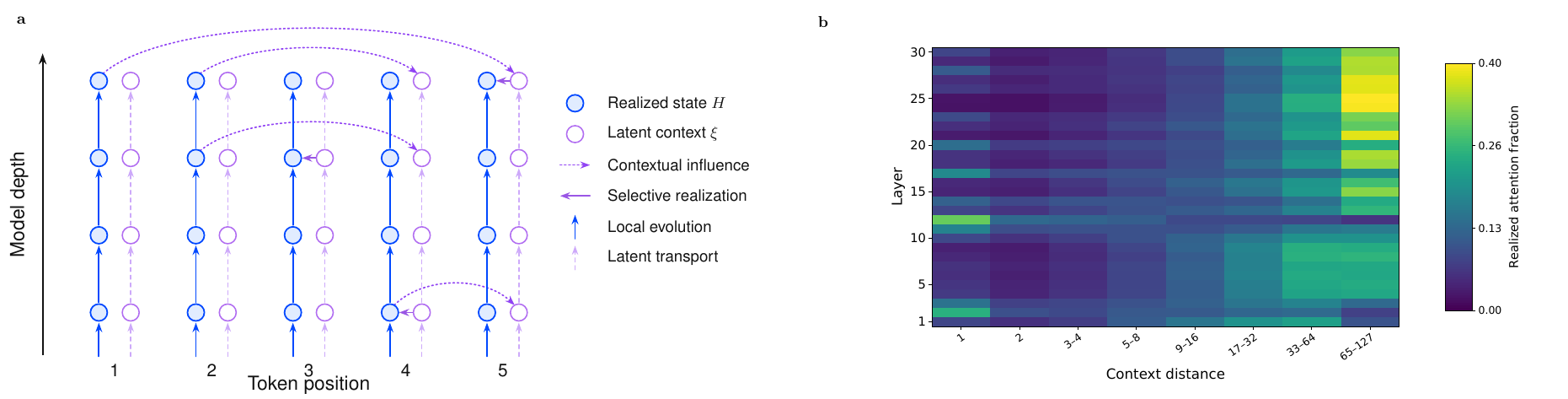}
\caption{\textbf{A Transformer architecture with finite-transport latent contextual influence and selective realization.}
\textbf{a, Layerwise causal-fate construction.} At each layer, every token maintains a realized state and a latent contextual state, whose sum forms the complete state. The complete Transformer map transports and transforms the latent contextual influence across depth, while a token-level realization rule determines when the accumulated influence enters the realized hidden state.
\textbf{b, Depth-dependent organization of realized contextual influence.} The heat map shows the fraction of realized cross-token attention mass across Transformer layers and source--target context-distance bins on the held-out WikiText-2 test blocks, normalized within each layer over the distance bins.}
\label{fig:transformer-causal-fate}
\end{figure*}

The architecture and the resulting organization of realized contextual attention are shown in Fig.~\ref{fig:transformer-causal-fate}.

The selective-realization Transformer achieved a perplexity of \(39.9138\), only \(3.27\%\) higher than that of the fully realized model, while realizing \(88.40\%\) of token--layer influence opportunities. On average, each token admitted accumulated contextual influence in \(26.52\) of the \(30\) layers, leaving \(3.48\) layers in which its realized component continued through token-local computation while contextual influence remained latent.
The active-layer ratio was 1: every layer contained at least one token for which cross-token influence was realized. Cross-token coordination therefore remained active throughout the network depth, although the participating token positions varied across layers.

This experiment evaluates functional viability rather than computational
efficiency. It quantifies how frequently contextual influence enters the realized
token state within the implemented causal-fate architecture. The modified computation graph directly implements the causal-fate
representation. It explicitly maintains, transports and selectively
realizes latent contextual influence while performing a non-trivial
language-modeling task. This construction establishes causal-fate dynamics as
a viable architectural principle for explicitly carrying and selectively realizing contextual influence.

\section{Discussion}\label{sec:discussion}

The formal construction shows how an unrealized consequence can remain part of
the dynamical state. Its aggregate latent displacement is transported and
transformed by the system map, and a realization rule determines what enters
the realized trajectory. When the relevant finite maps and the fully realized
reference can be specified in additive coordinates, the complete represented
state evolves exactly under the fully realized map. This is one exact
representation of causal-fate dynamics, not a claim that every system has the
same ontology or coordinate decomposition. The predictive-state and
aggregate-uniqueness results, including channel-level representation freedom,
are proved in Supplementary Note~1.

In the connectome-constrained \textit{Caenorhabditis elegans} model,
local-autonomy dynamics failed to reproduce the selected cross-neuronal
propagation structure, whereas selective realization of transported latent
influence preserved its principal form. This is a functionally non-trivial
witness within the adopted neural model and motivates a biological hypothesis:
unresolved inter-neuronal influence may be carried by distributed physical
processes whose later release, transformation or absorption contributes to
neural propagation. The present results do not identify such carriers in
living animals.

The BGP study provides observational support for the operational relevance of history. A bounded ordered
cross-observer history retained short-term predictive information beyond the
current local route. Within-sample controls distinguished event-slot encoding
from chronological content: age-free chronological content improved over random
content order in the primary comparison, but its advantage over a canonical
content inventory depended on the model and seed. The
front changed through matching, consumption and expiry rather than serving as
a static archive. This analysis does not establish a direct causal edge
between collectors and does not verify the additive finite-transport
equations. It identifies a finite, dynamically updated history representation that
retains additional predictive information in the tested routing models.

The Transformer study turns the same representation into an executable
architectural principle. Its computation graph explicitly implements
token-local evolution, transported latent contextual influence and selective
realization across depth. The result establishes architectural viability, not
computational efficiency, attention-cost reduction or improved predictive
performance.

The studies therefore progress through three distinct roles: a model-motivated scientific hypothesis in the neural system, an observational phenomenon compatible with future-relevant history in routing, and a direct executable construction in the Transformer. They do not jointly verify a universal physical mechanism. The exact finite-transport relation follows from the construction and its assumptions; empirical relevance depends on the chosen state, evolution maps and realization rule. Causal-fate dynamics provides a framework for formulating hypotheses about unfinished influence, examining its predictive signatures and designing systems that explicitly carry and selectively realize it.

\section{Methods}\label{sec:methods}

\subsection{Rule-defined chess illustration}

The chess illustration in Fig.~\ref{fig:chess-history-witness} is defined
entirely by the rules of chess. In panel a, the visible piece placement and
side to move are held fixed, while castling is permitted only in the history
in which neither the king nor the relevant rook has previously moved. In panel
b, the visible placement, side to move and move rights are held fixed, while
the position is shown on its first versus third occurrence; under the
threefold-repetition rule, the latter permits a draw claim. These constructions
follow Articles 3.8.2.1 and 9.2 of the FIDE Laws of Chess~\cite{fide2023laws}.

\subsection{\textit{C. elegans} neural-system experiment}

\subsubsection{Data, model and causal-fate construction}

The neural-system experiment used the public \textit{Caenorhabditis elegans}
signal-propagation atlas of Randi et al.~\cite{randi2023neural}, anatomical
connectivity through Worm Neuro Atlas~\cite{wormneuroatlas2023}, the adult and
developmental connectome resources represented in that atlas
~\cite{white1986structure,witvliet2021connectomes}, and its chemical-synapse
sign-prediction resources~\cite{fenyves2020synaptic}. The evaluated head
network contained 188 neurons, including directed chemical synapses and
electrical gap junctions. The fixed pairwise propagation object contained 168
stimulus neurons, 185 response neurons, 23,264 observed off-diagonal pairs,
23,180 valid kernel pairs and 1,141 high-confidence pairs. Each pair was
summarized by a signed-mean observation operator over a common 10-s response
window.

The continuous substrate followed the connectome-constrained graded-potential
model structure of Kunert, Shlizerman and Kutz~\cite{kunert2014low}, within
the same model family used for multistability studies
~\cite{kunertgraf2017multistability}. The fully coupled map retained
inter-neuronal dynamics; the local-autonomy map retained intrinsic neuronal
evolution, local gating and external stimulation while withholding newly
generated chemical-synaptic and electrical coupling influence. Connectome
preprocessing, continuous equations, readout fitting and parameterization are
given in Supplementary Note~2.

The selective trajectory used the finite-transport recurrence defined in the
Results and implemented over the fully coupled and local maps. It maintained a
realized state and latent residual, computed candidate non-local influence from
the complete state, released the selected component and retained the remainder
as the next latent residual. The exact represented-state closure and numerical
checks are given in Supplementary Note~2.

The 168 stimulus neurons were split once into 118 selection stimuli and 50
evaluation stimuli using seed 188. The continuous model, empirical masks and
readout were fixed before this split. The realization threshold and interval
were selected using only the 118 selection stimuli and then frozen before the
50 evaluation stimuli were scored. The complete operating-point protocol is
reported in Supplementary Note~2.

\subsubsection{Propagation-function evaluation}

For predicted propagation matrix \(P\), target matrix \(Y\) and Boolean pair
mask \(M\), valid evaluation entries were
\[
\mathcal V=\{(i,j):M_{ij}=1,\;P_{ij}\ {\rm is\ finite},\;Y_{ij}\ {\rm is\ finite}\}.
\]
The primary propagation metric was relative error,
\[
E_{\rm prop}(P,Y;M)=
\frac{\left(\sum_{(i,j)\in\mathcal V}(P_{ij}-Y_{ij})^2\right)^{1/2}}
{\left(\sum_{(i,j)\in\mathcal V}Y_{ij}^2\right)^{1/2}}.
\]
Pearson correlation and sign agreement were computed on the same valid mask;
the local-autonomy comparison used the fully realized propagation matrix as
the target. Full calibration, threshold criteria and additional metrics are
given in Supplementary Note~2.

\subsection{BGP routing history-state identification}
\label{subsec:bgp-history-state}

\subsubsection{Data sources and target}

We used public MRT data from the University of Oregon Route Views Project and
the RIPE NCC Routing Information Service~\cite{routeviews2025,riperis2025}.
The collectors were RouteViews Chile and RIPE RIS \texttt{rrc06}, and the fixed
window ran from 6 July 2025 00:00 UTC to 7 July 2025 00:00 UTC. Each collector's
starting routing-information base initialized its local peer--prefix route
state, after which announcements and withdrawals were processed in UTC event-
time order under BGP-4 semantics~\cite{rekhter2006bgp}.

For each prediction direction, one collector was treated as local and the
other as remote. The unit of analysis was a prefix--observer-peer pair within
a convergence episode. A prefix-specific episode consisted of consecutive
observed event groups separated by no more than 300 s; a larger gap began a
new episode. Prefixes were selected from those active at both collectors, and
one fixed anchor peer was selected per collector before model fitting. The
binary target was whether the local peer--prefix route changed strictly after
the sample time and within the next 30 s. The held-out test set contained
13,630 samples from 81 complete episodes and 14 prefixes. The complete
selection and reconstruction rules are reported in Supplementary Note~3.

\subsubsection{Representations and online state update}

The strong current-state representation included collector and observer
identity, prefix and peer identity, current route presence, path length and
identity, origin, community count, latest local update type and age, route-
change status and time of day. The ordinary local-lag comparison added local
announcement and withdrawal counts over 30-s and 120-s windows.

The ordered propagation-front representation augmented the strong current
state with the types, ages and path-identity buckets of the five newest
unmatched remote updates, plus fixed-dimensional unmatched-queue summaries and
cross-observer mismatch indicators. A local event consumed the oldest remote
occurrence with the same event type and announcement path. Entries older than
300 s expired. All events at a shared timestamp were added before matching,
and the five-slot readout was constructed after matching and expiry. Queue
summaries described all eligible entries, whereas only the five newest entries
were represented individually.

The age-preserving encoding control permuted valid event tuples only within
each sample, retaining the same content, ages, padding and current-state
features. A separate order comparison omitted all per-event ages from both
the chronological and shuffled content representations. A canonical content
inventory provided a deterministic unordered control. All paired controls
shared the same unordered summaries; none transferred histories between
samples. Complete rules and feature definitions are given in the
Supplementary Note~3.

\subsubsection{Models, chronological split and uncertainty}

We evaluated Logistic regression and HistGradientBoosting. Hyperparameters and
data-dependent preprocessing were selected on validation log loss only and
then frozen. Samples used a chronological 60/20/20 split, with complete
convergence episodes assigned to one split and boundary episodes excluded.
All features were constructed online from timestamps no later than the sample
time, and the target used only strictly later local route changes.

The primary metric was held-out log loss. Uncertainty was estimated by
resampling continuous 30-min test blocks with replacement 1,000 times and
reporting the empirical 2.5th and 97.5th percentiles as a 95\% interval. The
full model specification, seed robustness and heterogeneity analyses are
reported in Supplementary Note~3.

\subsection{Causal-fate Transformer experiment}

\subsubsection{Model and evaluation data}

The Transformer experiment used the publicly available SmolLM2-135M model
~\cite{allal2025smollm2}, a decoder-only causal Transformer with 30 layers,
hidden size 576, 9 attention heads and 3 key--value heads. Pretrained weights
were kept fixed. Evaluation used WikiText-2~\cite{merity2016pointer},
tokenized into contiguous length-128 blocks with 33 validation blocks and 65
test blocks, yielding 8,255 effective test prediction positions after the
next-token shift. Validation data were used only for realization-threshold
selection; all reported test metrics used the frozen model and rule.

\subsubsection{Layer construction and metrics}

Transformer depth was the evolution axis. At each layer, the full pretrained
map retained causal self-attention, while a token-local map retained the
residual stream, normalization, position-wise computation and same-position
attention path. The complete state was propagated through the full map, the
realized state through the token-local map, and the difference was carried as
latent contextual influence until the token-level realization rule admitted
it. The complete layer recurrence, vector release rule and threshold-selection
protocol are given in Supplementary Note~4.

We evaluated shifted next-token cross entropy and perplexity, together with
the token--layer realization ratio and active-layer ratio. These metrics
describe the implemented realization process; they do not measure attention
cost or computational speed.

\backmatter

\section*{Data availability}
The public \textit{C. elegans} signal-propagation measurements used in this study are from Randi et al.~\cite{randi2023neural}, with machine-readable data available through the associated OSF repository (\url{https://doi.org/10.17605/OSF.IO/E2SYT}) and the online atlas at \url{https://funconn.princeton.edu}. The implementation used the public Worm Neuro Atlas repository (\url{https://github.com/francescorandi/wormneuroatlas.git}) at commit \texttt{b2e13d88b670efcb3438aeacba2ad4bd6c383933}. The anatomical connectome matrices used by Worm Neuro Atlas combine the White et al. adult connectome and the Witvliet et al. developmental connectomes under its default settings~\cite{white1986structure,witvliet2021connectomes}; chemical-synapse sign predictions were taken from the Fenyves et al. data distributed with Worm Neuro Atlas~\cite{fenyves2020synaptic}.

The BGP analysis used public MRT archives from the University of Oregon Route Views Project (\url{https://archive.routeviews.org/}) and the RIPE NCC Routing Information Service (\url{https://data.ris.ripe.net/})~\cite{routeviews2025,riperis2025}. The analysed collectors were RouteViews Chile (\texttt{route-views.chile}) and RIPE RIS \texttt{rrc06}, over 6 July 2025 00:00 UTC to 7 July 2025 00:00 UTC, with each collector's starting routing-information base and update stream. The large public source archives are not redistributed with the manuscript; study-generated processed tables and figure-source records are not deposited in a separate external archive.

The Transformer experiment uses the public WikiText dataset page at \url{https://huggingface.co/datasets/Salesforce/wikitext}, configuration \texttt{wikitext-2-raw-v1}~\cite{merity2016pointer}, and the public SmolLM2-135M model page at \url{https://huggingface.co/HuggingFaceTB/SmolLM2-135M}, revision \texttt{93efa2f097d58c2a74874c7e644dbc9b0cee75a2}~\cite{allal2025smollm2}. Study-generated result objects used by the manuscript are not deposited in a separate external data archive.

\section*{Code availability}
The code for the \textit{C. elegans}, BGP and Transformer experiments is maintained at \url{https://github.com/Hotaru366/causal-fate-code}. It contains source code, configurations, tests and instructions for obtaining inputs from their original providers and reproducing the experiments. Third-party raw data and pretrained model weights are not redistributed in the code repository.

\section*{Reproducibility statement}
The external resources used in this study are publicly available from their original providers. The neural experiment fixes the Worm Neuro Atlas source revision, kernel-window rule, connectome and readout, followed by realization-policy selection on 118 stimuli and evaluation on the remaining 50. This stimulus split is not independent validation of the calibrated neural substrate. The BGP analysis fixes the public RouteViews Chile and RIPE RIS \texttt{rrc06} collectors, the UTC window from 6 July 2025 00:00 to 7 July 2025 00:00, a chronological 60/20/20 split with whole convergence episodes, and a frozen evaluation protocol. The Transformer experiment uses the public SmolLM2-135M model revision \texttt{93efa2f097d58c2a74874c7e644dbc9b0cee75a2} with the WikiText \texttt{wikitext-2-raw-v1} configuration. Saved result tables, rather than refitted models, drive the manuscript numbers and figures; source code and execution instructions are maintained in the repository identified in the Code availability statement. Study-generated result objects are not deposited in a separate public archive.

\section*{Acknowledgements}
This work was supported in part by the National Natural Science Foundation of China under Grant 52307145, in part by the Shenzhen Basic Research Fund (Natural Science Foundation) under Grants QNXMB20250701091813018 and JCYJ20240813113532042, and in part by the Shenzhen Research Institute of Big Data (SRIBD) under Grant J00220250001.

\bibliography{references}

\end{document}


\title[Supplementary Information]{Supplementary Information: Causal-fate dynamics of unrealized influence}

\author[1]{\fnm{Yiwei} \sur{Liu}}\email{yiweiliu1@link.cuhk.edu.cn}
\author*[2]{\fnm{Luwei} \sur{Yang}}\email{yangluwei@sribd.cn}
\author*[1,2]{\fnm{Shunbo} \sur{Lei}}\email{leishunbo@cuhk.edu.cn}

\affil[1]{\orgdiv{School of Science and Engineering}, \orgname{The Chinese University of Hong Kong, Shenzhen}, \orgaddress{\state{Guangdong}, \postcode{518172}, \country{China}}}
\affil[2]{\orgname{Shenzhen Research Institute of Big Data (SRIBD)}, \orgaddress{\city{Shenzhen}, \postcode{518172}, \country{China}}}

\maketitle

\newcommand{\CelegansNeuronCount}{188}
\newcommand{\CelegansGChem}{0.14}
\newcommand{\CelegansGGap}{0.02}
\newcommand{\CelegansGLeak}{0.7}
\newcommand{\CelegansStimulusAmplitude}{1.0}
\newcommand{\CelegansVHalf}{-26.0}
\newcommand{\CelegansKSlope}{4.0}
\newcommand{\CelegansCapacitance}{1.0}
\newcommand{\CelegansRestingPotential}{-35.0}
\newcommand{\CelegansSynapseRiseRate}{2.0}
\newcommand{\CelegansSynapseDecayRate}{1.0}
\newcommand{\CelegansExcReversal}{0.0}
\newcommand{\CelegansInhReversal}{-70.0}
\newcommand{\CelegansNeutralReversal}{-35.0}
\newcommand{\CelegansProductionDt}{0.0125}
\newcommand{\CelegansSimulationDuration}{30.0}
\newcommand{\CelegansStimulusSteps}{32}
\newcommand{\CelegansStimulusDuration}{0.4}
\newcommand{\CelegansDecisionInterval}{0.1}
\newcommand{\CelegansDecisionSteps}{8}
\newcommand{\CelegansProductionIntegrator}{rk4}
\newcommand{\CelegansGateThreshold}{0.01452580363706288}
\newcommand{\CelegansBlockThreshold}{0.0}
\newcommand{\CelegansGateMode}{whole}
\newcommand{\CelegansReadoutRidge}{0.02}
\newcommand{\CelegansReadoutCoefficientCount}{188}
\newcommand{\CelegansReadoutNonzeroCount}{151}
\newcommand{\CelegansReadoutIntercept}{0.0}
\newcommand{\CelegansValidationStimulusCount}{118}
\newcommand{\CelegansHeldoutStimulusCount}{50}

\newcommand{\BGPLogGain}{1.97\%}
\newcommand{\BGPLogCI}{[0.23\%, 3.37\%]}
\newcommand{\BGPLogTupleGain}{2.14\%}
\newcommand{\BGPLogTupleCI}{[1.75\%, 2.62\%]}
\newcommand{\BGPLogContentGain}{2.58\%}
\newcommand{\BGPLogContentCI}{[2.17\%, 3.09\%]}
\newcommand{\BGPLogInventoryGain}{0.96\%}
\newcommand{\BGPLogInventoryCI}{[0.41\%, 1.67\%]}
\newcommand{\BGPLogTimePositive}{7}
\newcommand{\BGPLogPrefixPositive}{9}
\newcommand{\BGPLogCollectorPositive}{1}
\newcommand{\BGPHgbGain}{9.54\%}
\newcommand{\BGPHgbCI}{[7.18\%, 11.45\%]}
\newcommand{\BGPHgbTupleGain}{2.27\%}
\newcommand{\BGPHgbTupleCI}{[1.50\%, 3.26\%]}
\newcommand{\BGPHgbContentGain}{1.20\%}
\newcommand{\BGPHgbContentCI}{[0.36\%, 2.25\%]}
\newcommand{\BGPHgbInventoryGain}{-0.32\%}
\newcommand{\BGPHgbInventoryCI}{[-0.99\%, 0.37\%]}
\newcommand{\BGPHgbTimePositive}{10}
\newcommand{\BGPHgbPrefixPositive}{11}
\newcommand{\BGPHgbCollectorPositive}{2}
\newcommand{\BGPPairCount}{100}
\newcommand{\BGPExactPairCount}{0}
\newcommand{\BGPPairProbabilityTolerance}{0.0130}

\section*{Supplementary Note 1: Predictive-state necessity and latent-coordinate identifiability}

\subsection*{Predictive-state necessity}

Let \(\mathfrak H_t\) denote the set of admissible system histories up to step
\(t\). A history \(\mathcal H_t\in\mathfrak H_t\) includes the initial
condition, past exogenous inputs, past realization decisions and all resulting
system states up to the current step. Let
\begin{equation}
\chi_t(\mathcal H_t)=x_t
\label{eq:supp-history-realized-state}
\end{equation}
denote the realized state produced by that history.

Let \(\pi_{t:}\) denote a fixed future protocol comprising the future finite
maps, local maps, exogenous inputs and realization rules. For an integer
\(h\geq1\), let
\begin{equation}
\mathcal P_{t,h}(\mathcal H_t;\pi_{t:})
\label{eq:supp-future-prediction}
\end{equation}
denote the realized state reached at step \(t+h\) when the system continues
from history \(\mathcal H_t\) under \(\pi_{t:}\).

The realized state \(x_t\) is predictively sufficient if, for every
\(h\geq1\) and every admissible future protocol \(\pi_{t:}\), there exists a
map \(\mathcal G_{t,h}^{\pi}\) such that
\begin{equation}
\mathcal P_{t,h}(\mathcal H_t;\pi_{t:})
=
\mathcal G_{t,h}^{\pi}\!\left(\chi_t(\mathcal H_t)\right)
\label{eq:supp-predictive-sufficiency}
\end{equation}
for every admissible history \(\mathcal H_t\).

\begin{theorem}[Necessity of an augmented predictive state]
\label{thm:supp-necessity-augmented-state}
Suppose that there exist two admissible histories
\(\mathcal H_t^{A},\mathcal H_t^{B}\in\mathfrak H_t\), a common future
protocol \(\pi_{t:}\) and a horizon \(h\geq1\) such that
\begin{equation}
\chi_t(\mathcal H_t^{A})
=
\chi_t(\mathcal H_t^{B}),
\label{eq:supp-matched-realized-state}
\end{equation}
but
\begin{equation}
\mathcal P_{t,h}(\mathcal H_t^{A};\pi_{t:})
\neq
\mathcal P_{t,h}(\mathcal H_t^{B};\pi_{t:}).
\label{eq:supp-different-futures}
\end{equation}
Then the realized state \(x_t\) is not predictively sufficient.

Moreover, let
\begin{equation}
y_t=\mathcal Y_t(\mathcal H_t)
\end{equation}
be any exact Markovian state representation, where \(\mathcal Y_t\) maps each
admissible history to its representation and the future realized trajectory is
determined under every admissible future protocol. Then
\begin{equation}
\mathcal Y_t(\mathcal H_t^{A})
\neq
\mathcal Y_t(\mathcal H_t^{B}).
\label{eq:supp-augmented-states-differ}
\end{equation}
In particular, if \(y_t=\operatorname{col}(x_t,m_t)\) for an auxiliary state
\(m_t\), where \(\operatorname{col}\) denotes column concatenation,
Eq.~\eqref{eq:supp-matched-realized-state} implies
\begin{equation}
m_t^{A}\neq m_t^{B}.
\label{eq:supp-memory-must-differ}
\end{equation}
Thus every exact predictive representation must retain history-dependent
information that cannot be recovered from the realized state alone.
\end{theorem}

\begin{proof}
Assume, for contradiction, that \(x_t\) is predictively sufficient. Under the
common future protocol \(\pi_{t:}\) and horizon \(h\),
Eq.~\eqref{eq:supp-predictive-sufficiency} assigns both histories the same
future because their realized states agree by
Eq.~\eqref{eq:supp-matched-realized-state}. This contradicts
Eq.~\eqref{eq:supp-different-futures}, so \(x_t\) is not predictively
sufficient.

If an exact Markovian representation assigned the same state to the two
histories, the common future protocol would again produce the same future
realized trajectory, contradicting Eq.~\eqref{eq:supp-different-futures}.
The augmented states must therefore differ. When their realized components
are equal, that distinction must be carried by the auxiliary components,
which gives Eq.~\eqref{eq:supp-memory-must-differ}.
\end{proof}

The theorem concerns predictive information, not a unique coordinate system.
An invertible change of coordinates on an augmented state can relabel the
missing information, but it cannot eliminate the need to distinguish histories
with the same realized present and different futures. In causal-fate dynamics,
this distinction is represented by the latent state. Its additional structure
comes from finite transport: the latent coordinate is constrained by the same
finite dynamics that defines the fully realized reference rather than evolving
as an arbitrary memory variable.

\subsection*{Aggregate uniqueness and channel-level representation freedom}

Let \(\mathcal C\) be the set of channels, \(c\in\mathcal C\) a channel index,
\(\mathcal X\) the realized-state coordinate space and \(\mathcal U_c\) the
coordinate space of channel \(c\). Let \(B_c:\mathcal U_c\rightarrow\mathcal X\)
embed that channel into realized-state coordinates. Let \(\xi_{c,t}\) be its
latent component at step \(t\), let \(\xi_t\) be the column stack of all such
components and let \(\bar\xi_t\) be their aggregate latent displacement. Define
the block embedding operator
\begin{equation}
\mathcal B
=
\begin{bmatrix}
B_1 & \cdots & B_{|\mathcal C|}
\end{bmatrix},
\qquad
\bar\xi_t=\mathcal B\xi_t,
\label{eq:supp-block-latent-embedding}
\end{equation}
where
\(\xi_t=\operatorname{col}(\{\xi_{c,t}\}_{c\in\mathcal C})\).

\begin{proposition}[Aggregate uniqueness and channel-level representation freedom]
\label{prop:supp-aggregate-uniqueness}
Fix the fully realized maps, local maps, exogenous inputs, initial complete
state and realized trajectory. In additive realized-state coordinates, the
aggregate latent displacement compatible with exact finite closure is unique:
\begin{equation}
\bar\xi_t=x_t^{\mathrm F}-x_t.
\label{eq:supp-aggregate-uniqueness}
\end{equation}
where \(x_t^{\mathrm F}\) is the fully realized reference state at step \(t\).
The channel-level latent state \(\xi_t\) is unique if and only if
\begin{equation}
\ker\mathcal B=\{0\}.
\label{eq:supp-channel-injectivity}
\end{equation}
where \(\ker\mathcal B\) denotes the null space of \(\mathcal B\).
When \(\ker\mathcal B\neq\{0\}\), the complete state determines only the
equivalence class
\begin{equation}
[\xi_t]_{\mathcal B}
=
\left\{
\xi_t+\nu:\nu\in\ker\mathcal B
\right\}.
\label{eq:supp-latent-equivalence-class}
\end{equation}
Here \(\nu\) is an arbitrary null-space displacement.
\end{proposition}

\begin{proof}
Exact finite closure gives \(x_t+\bar\xi_t=x_t^{\mathrm F}\), which uniquely
fixes the aggregate displacement in Eq.~\eqref{eq:supp-aggregate-uniqueness}
once \(x_t^{\mathrm F}\) and \(x_t\) are fixed. At the channel level, two
latent states \(\xi_t\) and \(\xi_t'\) produce the same aggregate displacement
precisely when
\begin{equation}
\mathcal B(\xi_t-\xi_t')=0.
\end{equation}
Every compatible pair is therefore identical if and only if \(\mathcal B\) is
injective. Otherwise, all states differing by an element of
\(\ker\mathcal B\) produce the same aggregate displacement and the same
complete represented state.
\end{proof}

Aggregate uniqueness and channel-level identifiability are thus distinct.
The fully realized reference fixes the aggregate latent displacement, but
overlapping channel embeddings leave freedom in its channel decomposition.
Claims about a particular latent channel require injective embeddings,
additional structural constraints or independent observations capable of
resolving the equivalence class in
Eq.~\eqref{eq:supp-latent-equivalence-class}.

\section*{Supplementary Note 2: \textit{C. elegans} model and protocol}
\label{sec:supp-celegans}

\subsection*{Public data and fixed propagation object}

The neural-system experiment used public \textit{C. elegans} signal-propagation measurements from Randi et al.~\cite{randi2023neural} and anatomical connectivity through Worm Neuro Atlas~\cite{wormneuroatlas2023}. The anatomical connectome resources represented in the atlas include the classical adult wiring diagram and later developmental connectome reconstructions~\cite{white1986structure,witvliet2021connectomes}. Chemical-synapse signs followed the sign-prediction resources associated with the atlas~\cite{fenyves2020synaptic}. The continuous neural dynamics instantiate the connectome-constrained graded-potential model structure of Kunert, Shlizerman and Kutz~\cite{kunert2014low}, within the same connectome-dynamics model family later used for multistability studies~\cite{kunertgraf2017multistability}.

The fixed evaluation object is a pairwise stimulus--response propagation matrix derived from the public signal-propagation atlas. This cross-experiment summary of stimulus--response measurements defines the propagation function to be evaluated and fits the study-specific readout.

\begin{table}[h]
\centering
\small
\caption{Fixed \textit{C. elegans} data object. Here \(q\) denotes the multiple-testing-adjusted \(q\) value, and the empirical transform uses response amplitude \(a\) with fixed scaling constant \(s\).}
\label{tab:supp-celegans-data}
\begin{tabular}{p{0.42\textwidth}p{0.46\textwidth}}
\hline
Quantity & Value \\
\hline
Head neurons & 188 \\
Stimulus neurons with observations & 168 \\
Response neurons with observations & 185 \\
Observed off-diagonal pairs & 23,264 \\
Valid observed kernel pairs & 23,180 \\
High-confidence \(q<0.05\) pairs & 1,141 \\
Observation window & 10 s \\
Observation operator & signed mean \\
Empirical transform & \(\tanh(a/s)\), with \(s\) fixed from high-confidence responses \\
Formal split seed & 188 \\
Selection/evaluation stimulus counts & 118 / 50 \\
\hline
\end{tabular}
\end{table}

\subsection*{Connectome preprocessing}

Chemical and gap-junction matrices are first restricted to the 188 selected head neurons. Chemical self-edges and gap-junction self-edges are set to zero. Each matrix is transformed with \(\log(1+w)\), where \(w\) is its non-negative connection count, and normalized by incoming row sum with a \(10^{-12}\) floor. Gap-junction counts are symmetrized before normalization. Chemical reversal potentials are assigned by sign: excitatory \(0\) mV, inhibitory \(-70\) mV and complex or neutral \(-35\) mV.

\subsection*{Neural model and full/local decomposition}

For \(N=\CelegansNeuronCount\), the number of head neurons, the state for each stimulus trajectory is \(x=(v,s)\in\mathbb R^{2N}\), with membrane voltage \(v_i\) in mV-like model units and synaptic gate \(s_i\in[0,1]\). The initial state sets \(v_i=\CelegansRestingPotential\) and \(s_i\) to the steady state of the gate equation at resting voltage. A stimulus current is applied to the stimulated neuron at the start of each trajectory.

For target-neuron index \(i\) and network-neuron index \(j\), let \(C_{ij}\), \(G_{ij}\) and \(E_{ij}\) denote the normalized chemical, gap-junction and reversal-potential matrices, respectively, and let \(u_i\) denote external current. Let \(\phi_i\) be the voltage-dependent gate-activation function; \(v_{1/2}\) and \(k\) its midpoint and slope; \(\alpha\) and \(\beta\) its activation and decay rates; \(I_i^{\rm gap}\) and \(I_i^{\rm chem}\) the gap-junction and chemical currents; \(g_{\rm gap}\), \(g_{\rm chem}\) and \(g_{\rm leak}\) their respective coupling or leak conductances; \(v_{\rm rest}\) the resting voltage; and \(C_m\) the membrane capacitance. A dot denotes a derivative with respect to continuous time. The full vector field is
\begin{align}
\phi_i(v_i)&=\left[1+\exp\{-(v_i-v_{1/2})/k\}\right]^{-1},\\
\dot s_i&=\alpha \phi_i(v_i)(1-s_i)-\beta s_i,\\
I_i^{\rm gap}&=g_{\rm gap}\left(v_i\sum_jG_{ij}-\sum_jG_{ij}v_j\right),\\
I_i^{\rm chem}&=g_{\rm chem}\left(v_i\sum_jC_{ij}s_j-\sum_jC_{ij}E_{ij}s_j\right),\\
\dot v_i&=\frac{-g_{\rm leak}(v_i-v_{\rm rest})-I_i^{\rm gap}-I_i^{\rm chem}+u_i}{C_m}.
\end{align}
The local map uses the same membrane, gate and stimulus terms but sets cross-neuronal chemical and gap-junction currents to zero, \(I_i^{\rm gap}=I_i^{\rm chem}=0\). This local/full split defines the operational decomposition within the connectome-constrained model.

\subsection*{Selective-realization protocol}

Let \(\Phi_{\rm F}\) be the full nonlinear map over one realization interval and \(\Phi_{\rm L}\) the local map over the same interval. For state \(x_t\) at discrete step \(t\), latent remainder \(r_t\), candidate non-local influence \(h_t\) and released component \(q_t\), the selective trajectory was computed as
\begin{align}
h_t&=\Phi_{\rm F}(x_t+r_t)-\Phi_{\rm L}(x_t),\\
x_{t+1}&=\Phi_{\rm L}(x_t)+q_t,\\
r_{t+1}&=h_t-q_t.
\end{align}
The implementation uses finite displacement as the main recurrence. Realization statistics describe the implemented selective-realization policy. Numerical checks confirmed that the implemented recurrence matched the intended finite-displacement relation to recorded precision.

The production integrator was \CelegansProductionIntegrator{}, with step
\(\Delta t=\CelegansProductionDt\) s after the integration-refinement audit.
The simulation lasted \CelegansSimulationDuration{} s, with stimulation during the initial \CelegansStimulusDuration{} s and
readout over the first 10 s. The selected realization interval was
\(\CelegansDecisionInterval\) s (\CelegansDecisionSteps{} integration steps),
with whole-state realization threshold \(\CelegansGateThreshold\). The substrate parameters and readout remained fixed during integration refinement.

\subsection*{Readout, calibration and stimulus-stratified evaluation}

For each simulated stimulus trajectory the model records membrane voltage \(v_i(t)\) and synaptic gate \(s_i(t)\), but the propagation readout uses only voltage. Let \(t\) denote continuous simulation time and \(T\) the selected 10-s observation-window length. For response neuron \(i\) and stimulus \(j\), the raw model feature is
\[
a_{ij}
=
\frac{1}{T}
\int_0^T
\left[v_i^{(j)}(t)-v_{\rm rest}\right]\,dt,
\]
where \(v_i^{(j)}(t)\) is the voltage of response neuron \(i\) under stimulus \(j\), implemented by a trapezoidal average over the selected 10-s observation window. Synaptic gate variables affect \(a_{ij}\) through the dynamics but are not additional regression features. The transformed propagation entry is
\[
P_{ij}=w_i a_{ij},
\]
with one response-specific scalar \(w_i\geq0\), no intercept and no stimulus-specific coefficient. For a fixed continuous-model candidate, \(w_i\) is fit independently for each response neuron by the closed-form non-negative ridge problem
\[
\min_{w_i\geq0}
\sum_{j\in\mathcal J_i}
\left(w_i a_{ij}-y_{ij}\right)^2
+
\alpha n_i\left(\operatorname{var}(a_i)+10^{-12}\right)w_i^2,
\]
where \(\mathcal J_i\) is the calibration mask for response neuron \(i\), \(n_i=|\mathcal J_i|\), \(y_{ij}\) is the frozen empirical propagation entry, \(a_i=(a_{ij})_{j\in\mathcal J_i}\) is the vector of calibration features, \(\operatorname{var}\) is sample variance, and \(\alpha=\CelegansReadoutRidge\) is the ridge-penalty multiplier in this readout objective. The fitted readout has \CelegansReadoutCoefficientCount{} coefficients, \CelegansReadoutNonzeroCount{} non-zero coefficients and zero intercept.

Continuous parameters were selected by calibration against the frozen propagation object. For each candidate parameter set, the full model was simulated against the frozen empirical propagation object, the readout above was fit to the high-confidence empirical calibration object, and a weighted substrate-calibration score combined relative error, Pearson loss, sign loss and leading-mode loss. The full empirical propagation object was used to instantiate the fixed continuous substrate and the study-specific propagation readout.

A prespecified 118/50 stimulus split was subsequently used to select the realization operating point and to report a stimulus-stratified evaluation of the frozen selective trajectory. The split isolates operating-point selection from evaluation after the continuous model and readout have been fixed with the full propagation object. The selected operating point was required to satisfy prespecified propagation-quality criteria on the selection stimuli, after which the lowest realization frequency among qualifying candidates was chosen. The criteria were relative propagation error \(E_{\rm prop}\leq0.05\), Pearson correlation at least \(0.999\) and sign agreement at least \(0.99\). The realization threshold and interval were selected using only the 118 selection stimuli; the 50 evaluation stimuli were reserved for scoring after this operating point was frozen. Realization statistics describe the implemented realization policy.

\begin{table}[h]
\centering
\small
\caption{Stimulus-stratified neural metrics.}
\label{tab:supp-celegans-results}
\begin{tabular}{lll}
\hline
Metric & Selective & Local autonomy \\
\hline
Pair count & 364 & 364 \\
Relative propagation error & 0.03515 & 1.0 \\
Pearson correlation & 0.999999991 & not defined \\
Sign agreement & 1.0 & not defined \\
RMSE & 0.007795 & 0.22176 \\
\hline
\end{tabular}
\end{table}

\section*{Supplementary Note 3: BGP routing history-state identification}
\label{sec:supp-bgp}

\subsection*{Public data and MRT reconstruction}

The routing analysis used public MRT archives from the University of Oregon
Route Views Project and the RIPE NCC Routing Information Service
~\cite{routeviews2025,riperis2025}. The two observation points were RouteViews
Chile and RIPE RIS rrc06. The fixed window began at 6 July 2025 00:00 UTC and
ended at 7 July 2025 00:00 UTC. For each observer, the routing-information base
at the start of the window initialized the peer--prefix route state; all update
files covering the subsequent 24 h supplied announcements and withdrawals.
Announcements updated route presence, path, origin and community count, whereas
withdrawals removed the local route, following BGP-4 update semantics
~\cite{rekhter2006bgp}.

The normalized observational identity was collector, peer autonomous-system
number, peer address and prefix. Events were ordered by UTC MRT event time,
then by archive-file order, record order and within-record update order. Events
sharing a timestamp were processed as one observed group. A convergence episode
was a prefix-specific run of updates in which consecutive observed event groups
were separated by no more than 300 s; a larger gap began a new episode, a
conservative operational definition motivated by the extended nature of BGP
convergence~\cite{labovitz2000delayed}.

Data-integrity checks verified the required routing-table and BGP update record
types, including announcement and withdrawal fields, and scanned 5,568,057 MRT
records without a parse error. The stage used 52 starting routes for the
selected anchor-peer and prefix combinations and reconstructed 481,402 anchor
updates. Feature construction was online: at every sample, the latest timestamp
contributing to a feature was no later than the sample timestamp. A route-change
target was accepted only when its event time was strictly later than the sample
time.

\subsection*{Prefix and peer selection}

There were 119,612 prefixes active at both public collectors during the scan.
The frozen selection ranked prefixes by the smaller of their event counts at
the two collectors, then by total event count and a deterministic prefix
tie-break, and retained 50 shared active prefixes. Within each collector, a
single anchor peer was selected by maximum selected-prefix coverage, then event
count and a deterministic peer tie-break. The resulting fixed peers were
AS27678 at RouteViews Chile and AS25152 at rrc06; each covered 29 of the 50
selected prefixes in the update stream. This selection was completed before
model evaluation.

Complete episodes were assigned to the chronological split described below.
After boundary-episode exclusion, the retained objects were:

\begin{table}[h]
\centering
\small
\caption{Chronological BGP analysis objects. Positives are local peer--prefix
route changes within the subsequent 30 s.}
\label{tab:supp-bgp-split-counts}
\begin{tabular}{lrrrr}
\hline
Split & Samples & Positives & Complete episodes & Prefixes \\
\hline
Training & 56,941 & 41,963 & 208 & 16 \\
Validation & 9,696 & 7,377 & 63 & 10 \\
Held-out test & 13,630 & 10,412 & 81 & 14 \\
\hline
\end{tabular}
\end{table}

\subsection*{Unmatched-front state and feature definitions}

For each local collector--peer--prefix, the unmatched queue contains remote
updates that have not been consumed by a corresponding local observation.
An event fingerprint is its announcement/withdrawal type together with its full
AS path for announcements (the path is empty for withdrawals). At each observed
timestamp, events older than 300 s are removed. All remote additions in that
timestamp group are then inserted before local matching, so a simultaneous
matching pair is treated symmetrically at the two collectors. Each local event
consumes at most one occurrence: the oldest unmatched event with the same
fingerprint. Age exactly 300 s remains eligible; age greater than 300 s expires.
The queues are cleared at an episode gap. These operations define observational
matching and a finite retention window, rather than measuring physical arrival
or proving that an expired event has no subsequent physical effect.

The candidate representation reads the five newest surviving queue entries,
newest first, with fixed empty padding when fewer than five remain. Matching
and expiry therefore remove events from both the queue summaries and the
five-slot view; an older eligible event can enter that view after a newer one
is consumed. The underlying queue can contain more than five entries. Queue
counts and balances summarize all retained entries; the implementation caps
that queue at 10,000 events. Ordinary remote update history is maintained
separately for the cumulative features and is not the source of the five slots.

\begin{table}[htbp]
\centering
\footnotesize
\caption{BGP representation components. All inputs are available at the sample timestamp.}
\label{tab:supp-bgp-features}
\begin{tabular}{p{0.25\textwidth}p{0.65\textwidth}}
\hline
Component & Definition and membership \\
\hline
Strong current state & Collector, remote observer and prefix identity; local peer AS; route presence, path length and identity bucket, origin, community count; latest local update type, age and change status; time-of-day sine and cosine. Shared by all extensions. \\
Local lags & Local update counts over 30 and 120 s, and announcement and withdrawal counts over 120 s. \\
Cumulative history & Remote update counts over 30 and 120 s, remote announcement and withdrawal counts over 120 s, and unmatched announcement--withdrawal balance. \\
Unordered history summary & Cumulative features plus unmatched count and distinct fingerprint count, current local--remote presence/path/origin mismatch, remote path length and identity bucket. Shared by all five-slot and content controls. \\
Five-event front & Type, age and path-identity bucket of the five newest unmatched events after consumption and expiry. Empty slots use type 0, age 3,600 s and path bucket 0. \\
Interpretable history summary & Unordered summary plus oldest/latest unmatched ages, announcement--withdrawal transitions within the queue, and latest ordinary remote-event type and age. \\
Queue-removed summary & Interpretable summary excluding unmatched counts, balance, distinct-event count and oldest/latest unmatched ages; it retains the transition count and other remote summaries. \\
\hline
\end{tabular}
\end{table}

\subsection*{Within-sample content and order controls}

Two paired control designs distinguish slot encoding from chronological content.
The first permutes the valid (type, age, path-identity) tuples \emph{within each
sample}, preserving that sample's complete tuple multiset, padding positions,
current state, shared summaries and target. No historical field is transferred
between samples. Since tuple ages still reveal chronology, this comparison
measures sensitivity to slot encoding and is not interpreted as independent
order identification.

The second design excludes all five per-event age fields from both sides.
The chronological-content representation retains the ordered (type, path-identity)
pairs; its shuffled counterpart independently permutes those pairs within the
same sample. Both use the identical strong current state and unordered summary,
with no queue-transition feature or other explicit order summary. A deterministic
inventory control sorts the same content pairs by type and then path-identity
bucket, without using time. All three preserve the same per-sample content
multiset, but only the chronological representation explicitly supplies the
observed sequence. These age-free comparisons test the predictive value of
chronological content arrangement conditional on the shared features. Their
performance also depends on the model family and representation; they are not
an estimator-independent test of conditional independence. Empty and single-event
fronts are unchanged by permutation.

The permutation seed is fixed before fitting, and the same construction is
applied separately to each sample in training, validation and test. The primary
seed is 1729; 2718 and 31415 are the robustness seeds. These corrections and
controls were fixed before fitting the revised analysis, after inspection of
the earlier analysis; they are not a prospective independent replication.

\subsection*{Model specification}

The model families were Logistic regression and histogram-based gradient
boosting (HGB). Collector, remote-observer and prefix fields were imputed by
their most frequent training value and one-hot encoded with unknown levels
ignored. All other fields were median-imputed and standardized using training
data. AS-path identity uses a deterministic integer hash bucket (2,048 nonempty
buckets, with 0 reserved for the empty path). These bucket codes enter the
numerical branch in the retained preprocessing; their numerical spacing is not
a measure of path similarity. This preprocessing is shared across the matched
comparisons.

Logistic regression used L-BFGS, at most 400 iterations and the validation grid
$C\in\{0.1,1.0\}$. HGB used learning rate 0.08, 80 iterations, minimum leaf
size 40, $L_2$ penalty 1.0 and a maximum-leaf grid of 15 or 31. Each
representation's hyperparameters were selected by validation log loss, after
which training and validation were combined for final fitting. The complete
chosen parameters, validation scores and held-out metrics are provided in the
machine-readable Supplementary Data. No test result was used to choose the
representation, grid or correction protocol.

\subsection*{Chronological split and uncertainty}

The original 24-h window and chronological 60/20/20 split were retained, with
boundaries at 6 July 2025 14:24 and 19:12 UTC. Episodes crossing a boundary were
excluded as whole objects, removing 209,233 otherwise eligible rows. Sample
identities, targets, current-state features and split membership were checked
against the preceding reconstruction and were unchanged. All features use
observations no later than the sample time; targets use strictly later local
route changes. The final 120 s of the source window were omitted for complete
follow-up at both horizons.

For reference loss $L_{\rm reference}$ and candidate loss $L_{\rm candidate}$,
the relative gain is $G=1-L_{\rm candidate}/L_{\rm reference}$. Positive gain
means lower candidate log loss. Paired uncertainty estimates resample the ten
consecutive 30-min held-out time blocks with replacement. Primary intervals
use 1,000 repetitions and the empirical 2.5th and 97.5th percentiles; seed and
horizon diagnostics use 500 repetitions. Within each comparison both models
are evaluated on exactly the same resampled rows. The bootstrap conditions on
the fitted models and this held-out window.

\subsection*{Held-out comparisons and ablations}

The corrected five-event front gave relative gains of \BGPLogGain{} for Logistic
and \BGPHgbGain{} for HGB against the strong current-state model, with 95\%
intervals \BGPLogCI{} and \BGPHgbCI{}, respectively.
Table~\ref{tab:supp-bgp-ablations} reports every representation, including
negative gains. Table~\ref{tab:supp-bgp-bootstrap} separates the age-preserving
encoding comparison from the age-free content-order comparisons.

\begin{table}[htbp]
\centering
\footnotesize
\setlength{\tabcolsep}{3pt}
\caption{Held-out log loss and relative gain against the strong current-state model. All front slots are read from the eligible unmatched queue.}
\label{tab:supp-bgp-ablations}
\begin{tabular}{p{0.40\textwidth}rrrr}
\hline
Representation & Logistic loss & Gain & HGB loss & Gain \\
\hline
Current route without latest event & 0.461229 & -1.05\% & 0.383889 & -13.83\% \\
Strong current state & 0.456428 & 0.00\% & 0.337255 & 0.00\% \\
Ordinary local lags & 0.460030 & -0.79\% & 0.345884 & -2.56\% \\
Cumulative history summary & 0.446902 & 2.09\% & 0.332171 & 1.51\% \\
Interpretable history summary & 0.455556 & 0.19\% & 0.315268 & 6.52\% \\
Within-sample tuple shuffle (with ages) & 0.457188 & -0.17\% & 0.312173 & 7.44\% \\
Unordered history summary & 0.449808 & 1.45\% & 0.311887 & 7.52\% \\
Five newest unmatched events (with ages) & 0.447414 & 1.97\% & 0.305086 & 9.54\% \\
Chronological content (without ages) & 0.447934 & 1.86\% & 0.308312 & 8.58\% \\
Shuffled content (without ages) & 0.459780 & -0.73\% & 0.312071 & 7.47\% \\
Canonical content inventory (without ages) & 0.452298 & 0.90\% & 0.307315 & 8.88\% \\
Summary without unmatched queue & 0.453092 & 0.73\% & 0.312024 & 7.48\% \\
\hline
\end{tabular}
\end{table}

\begin{table}[htbp]
\centering
\footnotesize
\setlength{\tabcolsep}{3pt}
\caption{Paired continuous-time-block bootstrap comparisons (1,000 repetitions, ten 30-min blocks). Tuple shuffling retains event ages and is an encoding control; the content comparisons omit all per-event ages.}
\label{tab:supp-bgp-bootstrap}
\begin{tabular}{lp{0.39\textwidth}rl}
\hline
Model & Comparison & Gain & 95\% interval \\
\hline
Logistic & Five-event front vs current state & 1.97\% & [0.23\%, 3.37\%] \\
Logistic & Five-event front vs tuple shuffle & 2.14\% & [1.75\%, 2.62\%] \\
Logistic & Content order vs content shuffle & 2.58\% & [2.17\%, 3.09\%] \\
Logistic & Content order vs content inventory & 0.96\% & [0.41\%, 1.67\%] \\
Logistic & Five-event front vs unordered summary & 0.53\% & [-0.00\%, 1.04\%] \\
Logistic & Five-event front vs queue-removed summary & 1.25\% & [0.70\%, 1.77\%] \\
HGB & Five-event front vs current state & 9.54\% & [7.18\%, 11.45\%] \\
HGB & Five-event front vs tuple shuffle & 2.27\% & [1.50\%, 3.26\%] \\
HGB & Content order vs content shuffle & 1.20\% & [0.36\%, 2.25\%] \\
HGB & Content order vs content inventory & -0.32\% & [-0.99\%, 0.37\%] \\
HGB & Five-event front vs unordered summary & 2.18\% & [1.26\%, 2.90\%] \\
HGB & Five-event front vs queue-removed summary & 2.22\% & [0.63\%, 4.08\%] \\
\hline
\end{tabular}
\end{table}

Without per-event ages, chronological content gave gains of
\BGPLogContentGain{} and \BGPHgbContentGain{} over within-sample shuffled content,
with intervals \BGPLogContentCI{} and \BGPHgbContentCI{}. Against the canonical
content inventory, the corresponding gains were \BGPLogInventoryGain{} and
\BGPHgbInventoryGain{}, with intervals \BGPLogInventoryCI{} and
\BGPHgbInventoryCI{}. These controls delimit which part of the observed gain
can be attributed to chronological content under the tested models. The primary
HGB interval against the canonical inventory crossed zero; one robustness seed
also gave an interval crossing zero against shuffled content. Thus the evidence
for an order-specific increment depends on the representation and model.

\begin{figure}[htbp]
\centering
\includegraphics[width=\textwidth]{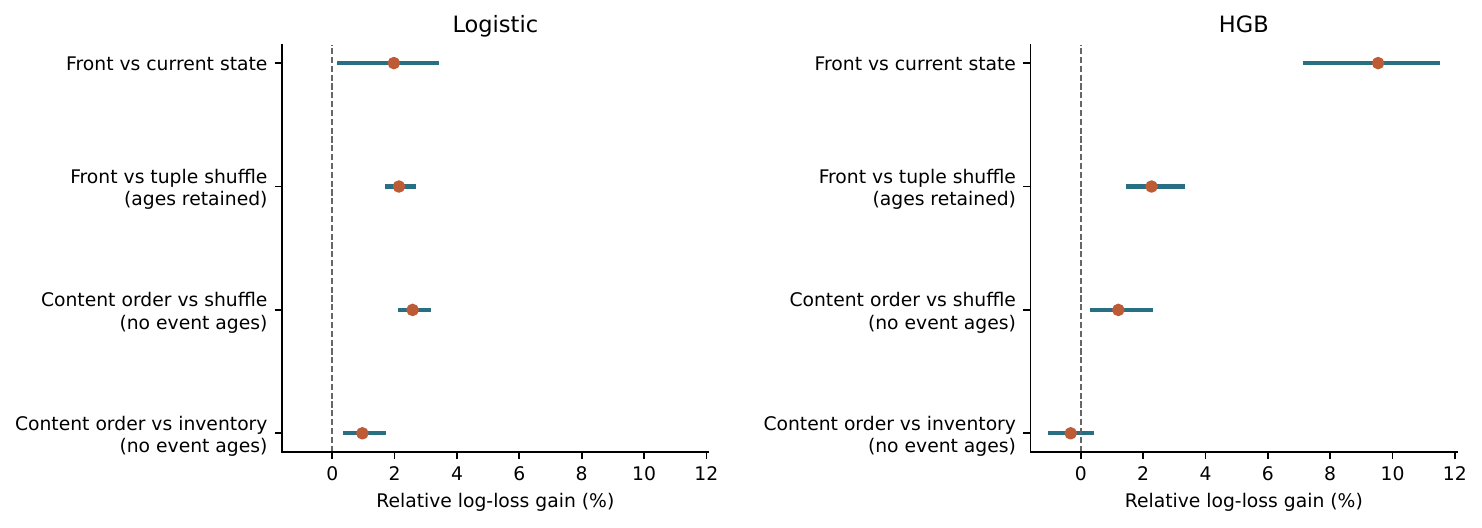}
\caption{Corrected unmatched-front comparisons. Points show relative held-out
log-loss gain and lines show paired 95\% block-bootstrap intervals. The
age-preserving tuple shuffle tests encoding; the two bottom comparisons omit
per-event ages and preserve each sample's event-content multiset.}
\label{fig:supp-bgp-controls}
\end{figure}

\subsection*{Time-block, collector and prefix heterogeneity}

Five-event-front gains were positive in \BGPLogTimePositive{} of ten time
blocks, \BGPLogCollectorPositive{} of two collector directions and
\BGPLogPrefixPositive{} of 14 prefixes for Logistic. The corresponding HGB
counts were \BGPHgbTimePositive{}, \BGPHgbCollectorPositive{} and
\BGPHgbPrefixPositive{}. All group results are retained below, including small
and negative-gain groups.

\begin{table}[htbp]
\centering
\scriptsize
\setlength{\tabcolsep}{3pt}
\caption{Five-event unmatched-front results by time block.}
\label{tab:supp-bgp-time-blocks}
\begin{tabular}{llrrrrr}
\hline
Model & Time block & Samples & Positives & Current & Front & Gain \\
\hline
Logistic & 1 & 312 & 244 & 0.4551 & 0.4541 & 0.21\% \\
Logistic & 2 & 1,225 & 1,034 & 0.3804 & 0.3805 & -0.02\% \\
Logistic & 3 & 1,313 & 1,086 & 0.3800 & 0.3563 & 6.25\% \\
Logistic & 4 & 1,388 & 1,063 & 0.5029 & 0.5125 & -1.90\% \\
Logistic & 5 & 1,600 & 1,258 & 0.4608 & 0.4434 & 3.78\% \\
Logistic & 6 & 1,764 & 1,332 & 0.4847 & 0.4671 & 3.63\% \\
Logistic & 7 & 1,609 & 1,337 & 0.3863 & 0.3749 & 2.94\% \\
Logistic & 8 & 1,495 & 1,157 & 0.4153 & 0.4165 & -0.28\% \\
Logistic & 9 & 1,607 & 1,131 & 0.4894 & 0.4733 & 3.30\% \\
Logistic & 10 & 1,317 & 770 & 0.6035 & 0.6010 & 0.42\% \\
HGB & 1 & 312 & 244 & 0.3472 & 0.3432 & 1.16\% \\
HGB & 2 & 1,225 & 1,034 & 0.2756 & 0.2751 & 0.18\% \\
HGB & 3 & 1,313 & 1,086 & 0.2781 & 0.2571 & 7.56\% \\
HGB & 4 & 1,388 & 1,063 & 0.3957 & 0.3635 & 8.13\% \\
HGB & 5 & 1,600 & 1,258 & 0.3449 & 0.3120 & 9.54\% \\
HGB & 6 & 1,764 & 1,332 & 0.3694 & 0.3219 & 12.87\% \\
HGB & 7 & 1,609 & 1,337 & 0.3360 & 0.3050 & 9.23\% \\
HGB & 8 & 1,495 & 1,157 & 0.2936 & 0.2617 & 10.88\% \\
HGB & 9 & 1,607 & 1,131 & 0.3245 & 0.2921 & 9.96\% \\
HGB & 10 & 1,317 & 770 & 0.4039 & 0.3445 & 14.71\% \\
\hline
\end{tabular}
\end{table}

\begin{table}[htbp]
\centering
\scriptsize
\setlength{\tabcolsep}{3pt}
\caption{Five-event unmatched-front results by collector.}
\label{tab:supp-bgp-collectors}
\begin{tabular}{llrrrrr}
\hline
Model & Collector & Samples & Positives & Current & Front & Gain \\
\hline
Logistic & RV Chile & 6,759 & 5,647 & 0.4192 & 0.3885 & 7.33\% \\
Logistic & RIS rrc06 & 6,871 & 4,765 & 0.4930 & 0.5053 & -2.50\% \\
HGB & RV Chile & 6,759 & 5,647 & 0.3231 & 0.2691 & 16.70\% \\
HGB & RIS rrc06 & 6,871 & 4,765 & 0.3512 & 0.3404 & 3.06\% \\
\hline
\end{tabular}
\end{table}

\begin{table}[htbp]
\centering
\scriptsize
\setlength{\tabcolsep}{3pt}
\caption{Five-event unmatched-front results by prefix.}
\label{tab:supp-bgp-prefixes}
\begin{tabular}{llrrrrr}
\hline
Model & Prefix & Samples & Positives & Current & Front & Gain \\
\hline
Logistic & 164.163.52.0/22 & 346 & 20 & 0.6804 & 0.7174 & -5.45\% \\
Logistic & 177.46.0.0/17 & 2,656 & 2,238 & 0.4155 & 0.5112 & -23.02\% \\
Logistic & 186.83.117.0/24 & 1,078 & 631 & 0.6721 & 0.6746 & -0.37\% \\
Logistic & 188.213.84.0/23 & 901 & 565 & 0.5778 & 0.5205 & 9.92\% \\
Logistic & 197.254.244.0/24 & 1,670 & 1,427 & 0.3243 & 0.2570 & 20.74\% \\
Logistic & 197.254.248.0/24 & 767 & 650 & 0.3406 & 0.2696 & 20.84\% \\
Logistic & 202.45.88.0/24 & 894 & 715 & 0.4339 & 0.3693 & 14.90\% \\
Logistic & 203.145.74.0/24 & 890 & 708 & 0.4430 & 0.3796 & 14.31\% \\
Logistic & 203.145.78.0/24 & 890 & 707 & 0.4440 & 0.3803 & 14.35\% \\
Logistic & 45.172.92.0/22 & 1,660 & 1,396 & 0.4415 & 0.4545 & -2.96\% \\
Logistic & 5.154.186.0/24 & 640 & 367 & 0.6879 & 0.7060 & -2.63\% \\
Logistic & 78.142.198.0/24 & 1,152 & 930 & 0.3860 & 0.3609 & 6.50\% \\
Logistic & 89.34.218.0/24 & 43 & 29 & 0.4545 & 0.3675 & 19.14\% \\
Logistic & 89.40.68.0/24 & 43 & 29 & 0.4546 & 0.3673 & 19.20\% \\
HGB & 164.163.52.0/22 & 346 & 20 & 0.1590 & 0.2624 & -65.09\% \\
HGB & 177.46.0.0/17 & 2,656 & 2,238 & 0.4353 & 0.4299 & 1.24\% \\
HGB & 186.83.117.0/24 & 1,078 & 631 & 0.4818 & 0.4816 & 0.04\% \\
HGB & 188.213.84.0/23 & 901 & 565 & 0.3660 & 0.3444 & 5.89\% \\
HGB & 197.254.244.0/24 & 1,670 & 1,427 & 0.1847 & 0.1563 & 15.41\% \\
HGB & 197.254.248.0/24 & 767 & 650 & 0.1996 & 0.1735 & 13.09\% \\
HGB & 202.45.88.0/24 & 894 & 715 & 0.3511 & 0.2364 & 32.68\% \\
HGB & 203.145.74.0/24 & 890 & 708 & 0.3560 & 0.2473 & 30.55\% \\
HGB & 203.145.78.0/24 & 890 & 707 & 0.3583 & 0.2468 & 31.11\% \\
HGB & 45.172.92.0/22 & 1,660 & 1,396 & 0.3434 & 0.3362 & 2.09\% \\
HGB & 5.154.186.0/24 & 640 & 367 & 0.4641 & 0.4345 & 6.37\% \\
HGB & 78.142.198.0/24 & 1,152 & 930 & 0.2065 & 0.1679 & 18.68\% \\
HGB & 89.34.218.0/24 & 43 & 29 & 0.2370 & 0.2506 & -5.73\% \\
HGB & 89.40.68.0/24 & 43 & 29 & 0.2369 & 0.2492 & -5.18\% \\
\hline
\end{tabular}
\end{table}

\subsection*{Seed robustness and alternative horizon}

The five-event front and both age-free content comparisons were refit under
all three seeds. Table~\ref{tab:supp-bgp-seeds} gives front-versus-current
results and Table~\ref{tab:supp-bgp-order-seeds} gives the order diagnostics.
The 120-s horizon and the interpretable history-summary representation are
reported alongside the primary 30-s front result in
Table~\ref{tab:supp-bgp-horizons}. These are secondary analyses, not replacement
criteria for selecting the headline result.

\begin{table}[htbp]
\centering
\scriptsize
\setlength{\tabcolsep}{3pt}
\caption{Seed robustness of the five-event front against the current state. Intervals use 500 block-bootstrap repetitions.}
\label{tab:supp-bgp-seeds}
\begin{tabular}{rllrrl}
\hline
Seed & Model & Reference & Reference loss & Candidate loss & Gain [95\% interval] \\
\hline
1729 & Logistic & Current & 0.456428 & 0.447414 & 1.97\% [0.28\%, 3.37\%] \\
2718 & Logistic & Current & 0.456428 & 0.447414 & 1.97\% [0.49\%, 3.38\%] \\
31415 & Logistic & Current & 0.456428 & 0.447414 & 1.97\% [0.56\%, 3.31\%] \\
1729 & HGB & Current & 0.337255 & 0.305086 & 9.54\% [7.02\%, 11.37\%] \\
2718 & HGB & Current & 0.342741 & 0.300615 & 12.29\% [8.83\%, 14.90\%] \\
31415 & HGB & Current & 0.333338 & 0.302459 & 9.26\% [5.49\%, 12.15\%] \\
\hline
\end{tabular}
\end{table}

\begin{table}[htbp]
\centering
\scriptsize
\setlength{\tabcolsep}{3pt}
\caption{Seed robustness of age-free chronological content. Reference is shuffled content or the canonical inventory. Intervals use 500 block-bootstrap repetitions.}
\label{tab:supp-bgp-order-seeds}
\begin{tabular}{rllrrl}
\hline
Seed & Model & Reference & Reference loss & Candidate loss & Gain [95\% interval] \\
\hline
1729 & Logistic & Shuffle & 0.459780 & 0.447934 & 2.58\% [2.18\%, 3.05\%] \\
1729 & Logistic & Inventory & 0.452298 & 0.447934 & 0.96\% [0.41\%, 1.68\%] \\
2718 & Logistic & Shuffle & 0.450398 & 0.447934 & 0.55\% [0.32\%, 0.83\%] \\
2718 & Logistic & Inventory & 0.452298 & 0.447934 & 0.96\% [0.40\%, 1.72\%] \\
31415 & Logistic & Shuffle & 0.455230 & 0.447934 & 1.60\% [1.31\%, 2.00\%] \\
31415 & Logistic & Inventory & 0.452298 & 0.447934 & 0.96\% [0.43\%, 1.76\%] \\
1729 & HGB & Shuffle & 0.312071 & 0.308312 & 1.20\% [0.39\%, 2.21\%] \\
1729 & HGB & Inventory & 0.307315 & 0.308312 & -0.32\% [-1.00\%, 0.33\%] \\
2718 & HGB & Shuffle & 0.301595 & 0.299213 & 0.79\% [0.16\%, 1.41\%] \\
2718 & HGB & Inventory & 0.300020 & 0.299213 & 0.27\% [-1.27\%, 1.67\%] \\
31415 & HGB & Shuffle & 0.299290 & 0.297513 & 0.59\% [-0.03\%, 1.14\%] \\
31415 & HGB & Inventory & 0.301048 & 0.297513 & 1.17\% [0.52\%, 1.73\%] \\
\hline
\end{tabular}
\end{table}

\begin{table}[htbp]
\centering
\scriptsize
\setlength{\tabcolsep}{3pt}
\caption{Primary and secondary prediction horizons. Intervals use 500 repetitions.}
\label{tab:supp-bgp-horizons}
\begin{tabular}{lr lrrl}
\hline
Model & Horizon (s) & Representation & Current & History & Gain [95\% interval] \\
\hline
Logistic & 30 & Five-event front & 0.456428 & 0.447414 & 1.97\% [0.28\%, 3.37\%] \\
Logistic & 30 & History summary & 0.456428 & 0.455556 & 0.19\% [-1.60\%, 1.61\%] \\
Logistic & 120 & Five-event front & 0.220203 & 0.206245 & 6.34\% [2.56\%, 11.10\%] \\
Logistic & 120 & History summary & 0.220203 & 0.208659 & 5.24\% [1.60\%, 9.42\%] \\
HGB & 30 & Five-event front & 0.337255 & 0.305086 & 9.54\% [7.02\%, 11.37\%] \\
HGB & 30 & History summary & 0.337255 & 0.315268 & 6.52\% [3.96\%, 8.77\%] \\
HGB & 120 & Five-event front & 0.168271 & 0.166209 & 1.22\% [-3.91\%, 5.90\%] \\
HGB & 120 & History summary & 0.168271 & 0.170287 & -1.20\% [-6.90\%, 4.50\%] \\
\hline
\end{tabular}
\end{table}

\subsection*{Matched observations and queue transitions}

The retained tolerance-matching procedure found \BGPPairCount{} pairs with
different 30-s outcomes; \BGPExactPairCount{} pairs matched every checked
current-state coordinate exactly. Pairs were required to share the discrete
local route features, occur within 300 s, differ in latest-local-update age by
at most 5 s, and differ in baseline HGB probability by at most 0.02. The
largest observed probability difference was \BGPPairProbabilityTolerance{}.
These outcome-selected pairs illustrate differing histories and outcomes;
they are not exact controlled-state interventions. Complete identities and
matching diagnostics are retained in Supplementary Data.

Table~\ref{tab:supp-bgp-cases} lists illustrative additions and removals in the
corrected unmatched queue. Each five-slot view is recomputed from that queue
after the update. Full prefix and collector identities are recorded alongside
the UTC intervals in the machine-readable case table. The examples describe
changes to the constructed state and do not independently identify the physical
cause of a later route change.

\begin{table}[htbp]
\centering
\scriptsize
\setlength{\tabcolsep}{3pt}
\caption{Illustrative unmatched-queue transitions; removal denotes consumption or expiry. Mismatch and future-change columns give before/after binary values.}
\label{tab:supp-bgp-cases}
\begin{tabular}{rllrrrrll}
\hline
Case & Type & UTC interval & Size pre & post & Net pre & post & Mismatch & Future \\
\hline
1 & Add & 00:00:01--00:00:04 & 1 & 2 & 1 & 2 & 1/1 & 1/1 \\
2 & Add & 00:00:04--00:00:31 & 2 & 3 & 2 & 3 & 1/1 & 1/1 \\
3 & Add & 00:00:31--00:00:35 & 3 & 4 & 3 & 4 & 1/1 & 1/1 \\
4 & Add & 00:00:35--00:01:01 & 4 & 5 & 4 & 5 & 1/1 & 1/1 \\
5 & Add & 00:01:01--00:01:04 & 5 & 6 & 5 & 6 & 1/1 & 1/1 \\
6 & Add & 00:01:04--00:01:31 & 6 & 7 & 6 & 7 & 1/1 & 1/1 \\
7 & Add & 00:01:31--00:01:34 & 7 & 8 & 7 & 8 & 1/1 & 1/1 \\
8 & Add & 00:01:34--00:02:01 & 8 & 9 & 8 & 9 & 1/1 & 1/1 \\
9 & Add & 00:02:01--00:02:04 & 9 & 10 & 9 & 10 & 1/1 & 1/1 \\
10 & Add & 00:02:04--00:02:22 & 10 & 11 & 10 & 11 & 1/1 & 1/1 \\
11 & Remove & 00:07:04--00:07:31 & 26 & 25 & 26 & 25 & 1/1 & 1/1 \\
12 & Remove & 00:07:31--00:07:34 & 25 & 24 & 25 & 24 & 1/1 & 1/1 \\
13 & Remove & 00:08:01--00:08:04 & 25 & 24 & 25 & 24 & 1/1 & 1/1 \\
14 & Remove & 00:08:36--00:09:01 & 24 & 22 & 24 & 22 & 1/1 & 1/1 \\
15 & Remove & 00:09:04--00:09:31 & 23 & 22 & 23 & 22 & 1/1 & 1/1 \\
16 & Remove & 00:09:31--00:09:36 & 22 & 21 & 22 & 21 & 1/1 & 1/1 \\
17 & Remove & 00:11:00--00:11:05 & 22 & 21 & 22 & 21 & 1/1 & 1/1 \\
18 & Remove & 00:11:37--00:13:00 & 22 & 19 & 22 & 19 & 1/1 & 1/1 \\
19 & Remove & 00:15:30--00:15:34 & 22 & 21 & 22 & 21 & 1/1 & 1/1 \\
20 & Remove & 00:16:04--00:16:31 & 21 & 20 & 21 & 20 & 1/1 & 1/1 \\
\hline
\end{tabular}
\end{table}

\clearpage

\clearpage
\section*{Supplementary Note 4: Transformer architecture and evaluation protocol}
\label{sec:supp-transformer}

\subsection*{Model and evaluation data}

The Transformer experiment used the public SmolLM2-135M model~\cite{allal2025smollm2}, a decoder-only causal Transformer with 30 layers, hidden size 576, intermediate size 1536, 9 attention heads, 3 key--value heads and 134,515,008 parameters. The public pretrained weights and tokenizer were used without fine-tuning.

Evaluation used WikiText-2 raw text~\cite{merity2016pointer}. Text was tokenized into contiguous length-128 blocks. The validation set contained 33 blocks and the test set contained 65 blocks, corresponding to 4,191 validation and 8,255 test next-token prediction positions after the language-modeling shift.

\subsection*{Full and token-local layer maps}

For layer index \(\ell\), the full layer map \(F_\ell\) is the original pretrained Transformer layer with causal self-attention. The token-local map \(L_\ell\) reuses the same layer parameters while preserving the residual stream, normalization modules, positional embeddings, same-position attention path and MLP under a token-local attention restriction. With \(L=30\) denoting the final layer index, final language-model logits are computed from the final realized state \(H_L\) only.

\subsection*{Finite-transport recurrence and token-level realization}

For layer \(\ell\), let \(H_\ell\) denote the realized token-state matrix, \(\xi_\ell\) the latent contextual state and \(Z_\ell=H_\ell+\xi_\ell\) the complete state. The candidate contextual influence is
\begin{align}
Z_\ell &= H_\ell+\xi_\ell,\\
h_\ell&=F_\ell(Z_\ell)-L_\ell(H_\ell).
\end{align}
The token gate returns a released influence \(\psi_\ell\), giving
\begin{align}
H_{\ell+1}&=L_\ell(H_\ell)+\psi_\ell,\\
\xi_{\ell+1}&=h_\ell-\psi_\ell.
\end{align}
Consequently,
\[
H_{\ell+1}+\xi_{\ell+1}=F_\ell(H_\ell+\xi_\ell)
\]
layer by layer.

For token position \(i\), the realization score \(s_{\ell,i}\) is token-level:
\[
s_{\ell,i}=
\frac{\lVert h_{\ell,i}\rVert_2}
{\max(\lVert L_\ell(H_\ell)_i\rVert_2,10^{-8})}.
\]
Here \(h_{\ell,i}\) is token \(i\)'s candidate contextual-influence vector, \(L_\ell(H_\ell)_i\) is its token-local output vector, \(\lVert\cdot\rVert_2\) is the Euclidean norm and \(10^{-8}\) is a numerical stability floor. If the score for token \(i\) exceeds the frozen threshold, the full vector \(h_{\ell,i}\) is released into the realized hidden state. Otherwise the full vector remains latent. Tokens in the same layer are gated independently, and the terminal latent state is not forced to release.

\subsection*{Validation selection and test evaluation}

The realization threshold was chosen only from validation data. The selection rule was the lowest non-trivial token--layer realization ratio whose validation perplexity was no more than \(1.06\) times the fully realized validation perplexity. The selected threshold was fixed before test evaluation. Test data were not used for threshold selection, and model weights were not updated.

\begin{table}[h]
\centering
\small
\caption{Transformer test metrics.}
\label{tab:supp-transformer-results}
\begin{tabular}{llll}
\hline
Trajectory & Perplexity & Cross entropy & Token--layer realization ratio \\
\hline
Fully realized & 38.6512 & 3.65458 & 1.0 \\
Token-local & 1315.24 & 7.18178 & 0.0 \\
Selective & 39.9138 & 3.68672 & 0.884022 \\
\hline
\end{tabular}
\end{table}

\begin{figure}[h]
\centering
\includegraphics[width=0.84\textwidth]{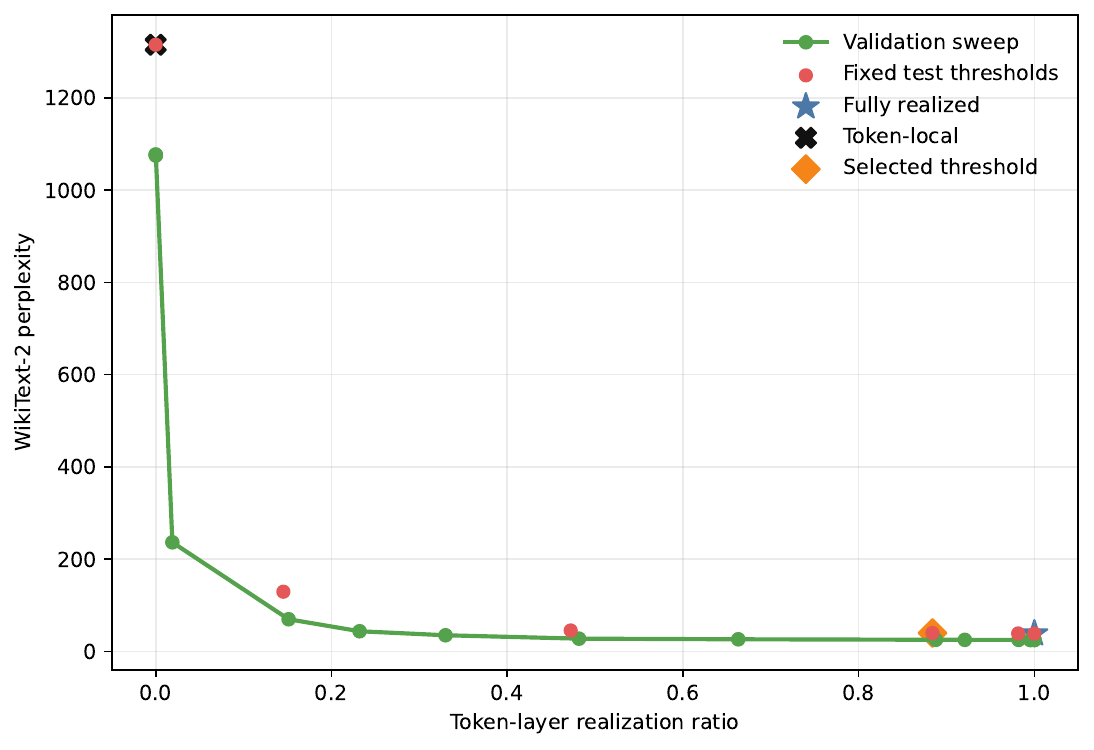}
\caption{Validation and test trade-off for the Transformer realization threshold.}
\label{fig:supp-transformer-threshold}
\end{figure}

\subsection*{Evaluation metrics}

Next-token cross entropy is computed from shifted logits and labels, summed over valid prediction positions and divided by the valid-token count. Perplexity is the exponential of this mean cross entropy. The token--layer realization ratio is the fraction of valid token--layer entries that release contextual influence. The active-layer ratio is the fraction of layers with at least one active token. Mean active layers per token equals the realization ratio times 30, and mean locally evolving layers per token is its complement.

Numerical checks confirmed that the custom layer loop and finite-transport recurrence matched the intended computation to recorded precision.

\section*{Code availability}
Source code and reproduction instructions for the three experiments are maintained at \url{https://github.com/Hotaru366/causal-fate-code}. Third-party inputs are obtained from their original providers rather than redistributed with the code.

\bibliography{references}